\documentclass[a4paper,11pt]{article}
\usepackage{jheppub,graphicx,floatflt}
\usepackage{amsmath,latexsym,amssymb,slashed}
\usepackage{enumerate}
\usepackage{mathrsfs}
\usepackage[colorlinks=true,%
            linkcolor=blue,%
            citecolor=blue,%
            urlcolor=blue]{hyperref}

\usepackage{dsfont}
\usepackage{multirow}
\newcommand\beal{\begin{align}}
\newcommand\nn{\nonumber}

\newcommand{\eq}[1]{\begin{equation}#1\end{equation}}
\newcommand{\spl}[1]{\begin{split}#1\end{split}}

\newcommand{\mcal}{\mathcal{M}}

\def\d{\text{d}}

\def\slashchar#1{\setbox0=\hbox{$#1$}           
\dimen0=\wd0                                 
\setbox1=\hbox{/} \dimen1=\wd1               
\ifdim\dimen0>\dimen1                        
\rlap{\hbox to \dimen0{\hfil/\hfil}}      
#1                                        
\else                                        
\rlap{\hbox to \dimen1{\hfil$#1$\hfil}}   
/                                         
\fi}

\newcommand{\id}{\mathds{1}}
\def\del {\partial}

\renewcommand{\i}{\ensuremath{\textnormal{i}}}

\def\f {{\rm \texttt{f}}}

\newcommand{\beq}{\begin{eqnarray}}
\newcommand{\eeq}{\end{eqnarray}}

\usepackage{xcolor}
\usepackage{color}

\newcommand{\cnote}[1]{}

\title{Towards Kaluza-Klein Dark Matter on Nilmanifolds}

\author[a,b]{David Andriot,}
\author[c]{Giacomo Cacciapaglia,}
\author[c,d]{Aldo Deandrea,}
\author[c,e]{Nicolas Deutschmann,}
\author[c]{and Dimitrios Tsimpis}

\affiliation[a]{Max-Planck-Institut f\"ur Gravitationsphysik, Albert-Einstein-Institut,
Am M{\"u}hlenberg 1, 14476 Potsdam-Golm, Germany}
\affiliation[b]{Institut f\"ur Mathematik, Humboldt-Universit\"at zu Berlin, IRIS-Adlershof, Zum Gro\ss en Windkanal 6, 12489 Berlin, Germany}
\affiliation[c]{Univ Lyon, Universit\'e Lyon 1, CNRS/IN2P3, IPNL, F-69622, Villeurbanne, France}
\affiliation[d]{Institut Universitaire de France, 103 boulevard Saint-Michel, 75005 Paris, France}
\affiliation[e]{Centre for Cosmology, Particle Physics and Phenomenology (CP3), Universit\'e catholique de Louvain, Chemin du Cyclotron 2, B-1348 Louvain-la-Neuve, Belgium}

\emailAdd{david.andriot@aei.mpg.de}
\emailAdd{g.cacciapaglia@ipnl.in2p3.fr}
\emailAdd{deandrea@ipnl.in2p3.fr}
\emailAdd{n.deutschmann@ipnl.in2p3.fr}
\emailAdd{tsimpis@ipnl.in2p3.fr}

\abstract{
We present a first study of the field spectrum on a class of negatively-curved compact spaces: nilmanifolds or twisted tori. This is a case where analytical results can be obtained, allowing to check numerical methods. We focus on the Kaluza-Klein expansion of a scalar field. The results are then applied to a toy model where a natural Dark Matter candidate arises as a stable massive state of the bulk scalar.
}

\begin{document}
\maketitle
\flushbottom
\setcounter{footnote}{0}
\renewcommand{\thefootnote}{\arabic{footnote}}
\setcounter{section}{0}

\section{Introduction}

Particle physics models are built on the basis of both experimental results and theoretical expectations. On the one hand one can start
from a well-defined high-energy theory and try to build a low-energy effective theory. A notable example is string theory which, in its best known phase,  lives in ten spacetime dimensions. Compactification of the six extra
dimensions provides us, in principle, with a systematic way to construct effective low-energy theories in four-dimensional spacetime. Another possibility consists of starting from the low-energy side and building intermediate theories that can then be linked to high-energy models. Although this second approach does not have the ambition to bring a complete solution all the way up to the Planck scale, it has the advantage of allowing the construction of new theories beyond the Standard Model, which may be able to tackle at least some
of the present problems. A popular example has been to consider extra compact spaces which can be large with respect
to the Planck scale, leading in some cases to measurable consequences at the TeV scale \cite{Antoniadis:1990ew,ArkaniHamed:1998rs,Antoniadis:1998ig,ArkaniHamed:1999dc}.
Compact manifolds studied in this setup are typically flat or positively-curved, stemming from the simple
examples of the torus or the sphere.

In recent years compact negatively-curved spaces (CNS) have been put forward in extra-dimensional models \cite{Kaloper:2000jb}, but they are much less studied than positively-curved compact spaces. They also appear in string theory \cite{Kehagias:2000dga,Orlando:2006cc,Silverstein:2007ac,Douglas:2010rt}, hence they can be connected to more fundamental theories.
The main motivation for the study of CNS is that some of their properties are extremely interesting for realistic model building.
Among them one can list:
\begin{itemize}
\item
Zero modes of the Dirac operator are typically present in the effective four dimensional theory \cite{Camporesi:1995fb}. This contrasts with the case of positive curvature, as for spherical orbifolds, where breaking of gauge invariance or background fields are needed,
see for example \cite{Maru:2009wu,Dohi:2010vc,Cacciapaglia:2016xty}.
\item
Explanation of the hierarchy between the Planck scale and the electroweak scale, thanks to their potentially large volume with respect to a
linear size only slightly larger than the new fundamental length scale (see for example \cite{Kaloper:2000jb}). In particular compact hyperbolic manifolds, which are special cases of CNS, have two length scales: $l_c$ linked to local
properties such as the curvature, and $l_G$, related to global properties such as the volume (which appears in the expression of the
effective Planck mass). Typically large $l_G$ values correspond to large genus. The volume grows exponentially with the ratio $l_G/l_c$, allowing a natural solution to the hierarchy problem  \cite{Orlando:2010kx}.
\item
Typical mass spectra for the Kaluza-Klein (KK) reduction of these spaces feature a mass gap similar to the one present in Randall-Sundrum models \cite{Randall:1999ee},
 therefore allowing for a solution to the hierarchy problem that does not require light KK modes \cite{Orlando:2010kx}.
\item
When the extra compact space is negatively-curved, rather than flat,  the dynamics of the very early universe alleviate standard
cosmological problems, allowing to account for the current homogeneity and flatness of the universe \cite{Starkman:2001xu,Chen:2003dca,Neupane:2003cs}.
\item
Finite volume hyperbolic manifolds of dimension greater than two have the remarkable
property of rigidity (Mostow's theorem \cite{mostow}): geometrical quantities as the length of its shortest closed geodesics, the volume,
etc. are topological invariants. In practice this means there are no more moduli once volume and curvature are fixed \cite{Nasri:2002rx}.
\end{itemize}
Different scenarios using these properties have been proposed for cosmology and inflation (see for example \cite{Greene:2010ch,Kim:2010fq}).
One striking point is the lack of detailed particle physics models based on orbifolds of CNS. The reason is the difficulty in
obtaining detailed predictions for the spectra of these models: for example the Laplacian eigenmodes of a generic compact hyperbolic manifold cannot be obtained in closed analytic form, thus the spectra must be sought by means of expensive numerical methods \cite{Cornish,Cornish2,Inoue}.

We shall therefore focus here
on the study of a specific case which allows analytical control of the spectra and their KK reductions: twisted tori, compact manifolds built as non-trivial
fibrations of tori over tori. They can be constructed from solvable (Lie) algebras and related groups, hence their
mathematical name of solvmanifolds. Nilpotent algebras and groups are subcases  thereof, from which one gets nilmanifolds. For the latter, the Ricci scalar is always negative; this also holds for most of the solvmanifolds. We have therefore at hand simple and calculable examples of CNS. Reviews on nil- and solvmanifolds can be found in \cite{Grana:2006kf, Andriot:2010ju, gong, bock}. Here, we will consider the unique three-dimensional nilmanifold ${\cal M}_3$ (apart from the torus $T^3$), built from the Heisenberg algebra, and we will determine the Laplacian scalar spectrum, both analytically and numerically.

The Laplacian spectrum on this manifold ${\cal M}_3$ has been determined in the mathematical literature \cite{Brezin, thang, Gordon, Schubert} for a certain canonical metric. We will present in Section \ref{sec:simple} these known mathematical results in a way more accessible to physicists, in the simple and explicit language of KK reductions. We then generalize the metric from the canonical one to include all extra metric parameters compatible with the Heisenberg algebra. Given the previous mathematical results, we easily obtain the Laplacian scalar spectrum now depending on all metric parameters. To our knowledge this result is new, and will be presented in Section \ref{sec:specgen}.

Twisted tori have played an important role in compactifications of string theory and supergravity down to four dimensions. Used as six-dimensional internal manifolds, they appear for instance in ten-dimensional vacua of type II supergravities. The first example of such a supersymmetric flux vacuum was found in \cite{Kachru:2002sk} on a four-dimensional Minkowski spacetime times ${\cal M}_3 \times T^3$. Many more have been found since, as e.g. in \cite{Grana:2006kf} (see \cite{Andriot:2015sia} for a review). ${\cal M}_3$ also appeared  as part of the internal manifold in the first example of axion monodromy inflation mechanism \cite{Silverstein:2008sg} (see \cite{Gur-Ari:2013sba, Andriot:2015aza} on the ten-dimensional completion of this cosmological model). A partial quantisation of closed strings on ${\cal M}_3$ appeared in \cite{Andriot:2012vb}. Since twisted tori are used as internal manifolds, it is important for string phenomenology to determine the corresponding low-energy effective theory in four dimensions. Four-dimensional gauged supergravities can be obtained by a truncation to a finite set of modes, as e.g. in \cite{Grana:2005ny,Caviezel:2008ik}. Whether this set corresponds to the light modes in a controlled low-energy approximation has not yet been settled. Contrary to reductions on pure Calabi-Yau manifolds or even torus, where the light modes are massless and the truncation of the KK tower is a good low-energy approximation, here the curvature induces additional energy scales that should be compared to the KK scale, and gives rise to massive light modes that complicate the truncation. The concrete KK scalar spectrum obtained in the present work should help clarify these issues.

The main phenomenological application we aim at is the construction of models containing a Dark Matter candidate in the form of a stable KK mode \cite{Servant:2002aq}. In particular, we want models where such candidates arise naturally, without the need to impose symmetries on the interaction terms localised on the singularities of the space, as in \cite{Cacciapaglia:2009pa}. Constructing models of Universal Extra Dimensions where all the fields propagate in the extra dimensions would require a detailed knowledge of the spectra for scalars, fermions and vectors. In this exploratory work we aim at constructing a toy model where a bulk scalar field provides a Dark Matter candidate, while the Standard Model particles (including the Higgs boson) live in four dimensions and are localised on a singular point of the internal space.

The paper is organised as follows. In Section~\ref{sec:explicit0} we present twisted tori and build our nilmanifold $\mcal_3$, defining in particular explicit coordinates together with discrete identifications that make the manifold compact; we also write down the most general left-invariant metric on $\mcal_3$. In Section~\ref{sec:KK} we analyse the KK spectrum of a scalar field for that metric, first in a simple case and then in the most general one.
Section~\ref{sec:num} contains a numerical procedure used to study spectra on CNS that we test against the analytical solutions obtained in the previous section.
Finally, in Section~\ref{sec:iso} we discuss how a Dark Matter candidate may arise in an orbifolded version of the twisted torus. We conclude in Section~\ref{sec:concl}. Further technical details can be found in the appendices.

\section{The three-dimensional nilmanifold}\label{sec:explicit0}

\subsection{From algebras to compact manifolds}\label{sec:explicit}

We consider here Lie algebras and Lie groups in one-to-one correspondence with the exponential map. Those are closely related to geometry, since any Lie group of dimension $d$ can be viewed as a $d$-dimensional manifold. To build a (compact) solvmanifold, or twisted torus, one divides a solvable Lie group by a lattice, namely a discrete subgroup that makes the manifold compact thanks to discrete identifications. Given a $d$-dimensional algebra generated by the vectors $\{Z_a,~a=1,\dots,d\}$ satisfying
\eq{\label{alg}[Z_b,Z_c]=f^a{}_{bc}Z_a~,}
with structure constants $f^a{}_{bc}=-f^a{}_{cb}$, the corresponding
$d$-dimensional manifold admits an orthonormal frame $\{e^a,~a=1,\dots,d\}$ obeying the Maurer--Cartan equation
\eq{\label{mc}\d e^a=-\frac12 f^a{}_{bc}e^b\wedge e^c~.}
These one-forms are globally defined on the manifold. One can see from \eqref{mc} that $f^a{}_{bc}$ is related to the spin connection. The Ricci tensor (in flat indices) can then be expressed in terms of the structure constants: for a unimodular Lie algebra, one gets
\eq{
{\cal R}_{cd}= \frac{1}{2} \left( - f^b{}_{ac} f^a{}_{bd} - \delta^{bg} \delta_{ah} f^h{}_{gc} f^a{}_{bd} + \frac{1}{2} \delta^{ah}\delta^{bj}\delta_{ci}\delta_{dg} f^i{}_{aj} f^g{}_{hb} \right)\; ,
\label{riccit}
}
with the flat, here Euclidian, metric $\delta_{ab}$; see e.g. \cite{Andriot:2015sia}. For a nilpotent algebra (subcase of solvable), and hence  for nilmanifolds, the first term is zero so the Ricci tensor is nowhere-vanishing and the corresponding Ricci scalar
\eq{
{\cal R}= -\frac{1}{4} \delta_{ad} \delta^{be} \delta^{cg} f^a{}_{bc} f^d{}_{eg}~, \label{Ricci}
}
is constant and strictly negative.

For $d=3$, in addition to the trivial abelian algebra leading  to the three-torus, there are three different
solvable algebras. Only one of them is nilpotent: it is the Heisenberg algebra, given by
\eq{\label{heis}[Z_1,Z_2]=- \f\, Z_3~,~~~[Z_1,Z_3]=[Z_2,Z_3]=0~,}
with $\f =-f^3{}_{12} \neq 0$, for which \eqref{mc} takes the form
\eq{\label{mcheis}
\d e^3= \f\, e^1\wedge e^2~;~~~\d e^1=0~;~~~\d e^2=0
~.}
The constant $\f$ thus has dimension of an inverse length. To build our $3$-dimensional nilmanifold $\mcal_3$, and to discuss in detail associated physical models, a set of explicit coordinates has to be selected. The following choice is consistent with \eqref{mcheis}
\eq{\label{a}e^1=r^1\d x^1~;~~~e^2=r^2\d x^2~;~~~e^3=r^3\left(\d x^3+N x^1\d x^2\right)~;~~~N=\frac{r^1r^2}{r^3}\f ~,}
for some constant (positive) ``radii'' $r^1$, $r^2$, $r^3$, and angular coordinates $x^m$. This is the most general solution up to redefining the coordinates.\footnote{Of course we could absorb the radii by rescaling the coordinates. However we have preferred to factor out the normalisation of the $x^m$'s so that we can impose integral identifications, see \eqref{heisids} below. Note also that fixing the value of $\f$ makes \eqref{heis} a single representative of a class of isomorphic algebras.}

For a compact three-dimensional $\mcal_3$, we impose the following discrete identifications (under a unit shift of the angles)
\eq{\label{heisids}
x^1\sim x^1+n^1~;~~~x^2\sim x^2+n^2~;~~~x^3\sim x^3+n^3- n^1N x^2~,~~~n^1,n^2,n^3\in\{ 0,1\}~,
}
which come from demanding that \eqref{heisids} should leave \eqref{a} invariant. Choosing different integer values for the $n^m$ is in general allowed, but corresponds to another lattice. We see from \eqref{heisids} that $\mcal_3$ is a twisted $S^1$ fibration over $T^2$, i.e. a twisted torus, with fiber coordinate  $x^3$ and base parameterised by $x^1$, $x^2$. The discrete identifications \eqref{heisids} correspond to the lattice action, making $\mcal_3$ a quotient of a nilpotent group by a discrete subgroup: a nilmanifold. Finally, for consistency we must require
\eq{\label{quant}N\in\mathbb{Z}^*~.}
The integrality of $N$ is discussed in Appendix \ref{sec:N}. Once the unit lattice in \eqref{heisids} is defined, one can verify that translations along the three directions with arbitrary $n^m \in \mathbb{Z}$ are also identified as in \eqref{heisids}. In addition, the lattice cell defined by a non unitary identification is equivalent to a unit cell with
\eq{{r'}^m = n^m r^m\,, \qquad N' = N \frac{n^1 n^2}{n^3} = \f \frac{{r'}^1 {r'}^2}{{r'}^3}~.}
From the results in Appendix \ref{sec:N}, it follows that $N' \in \mathbb{Z}^\ast$, thus implying that
\eq{\label{eq:n1n3} N \frac{n^1 n^2}{n^3} \in \mathbb{Z}^\ast~.}

To summarise, the space $\mcal_3$ is characterised by three independent radii $r^{m=1,2,3}$ and an integer $N$, related as in \eqref{a} to the structure constant or geometric flux $\f$. The coordinates $x^m$ chosen are angles ranging in $[0,1]$. In terms of physical dimensions, $x^m$ are dimensionless while the $r^m$ are a length, and $\f$ is the inverse of a length, i.e. an energy.

\subsection{General left-invariant metric on $\mcal_3$}
In vielbein basis the most general left-invariant metric for $\mcal_3$ reads
\eq{\label{heism}\d s^2= \delta_{ab}E^a E^b~,}
where $E^a$ are one-forms related to the orthonormal frame $e^a$ through
a constant $GL(3,\mathbb{R})$ transformation $L$
\eq{\label{e}E^a=\left(L^{-1}\right)^a{}_b e^b~.}
This description is redundant since we should mod out by the group $Aut$ of automorphisms of the Heisenberg algebra, i.e. by any $M\in Aut$ that leaves \eqref{mc} invariant. The $M$'s such that $m^a{}_b e^b$ satisfy \eqref{mc} are constrained by
\eq{m^a{}_d f^d{}_{ef} = f^a{}_{bc} m^b{}_e m^c{}_f~,}
condition that, for the Heisenberg algebra, is solved by a matrix of the form (see also \cite{Rezaei}):
\eq{
 M=\left(
\begin{array}{ccc}
m_{11}& m_{12} & 0\\
m_{21}& m_{22} & 0\\
m_{31} & m_{32} &m_{11}m_{22}-m_{12}m_{21}
\end{array}\right)\in Aut \ , \ m_{11}m_{22}-m_{12}m_{21} \neq 0
~.
}
We want to mod out by such elements, i.e. we want to consider in \eqref{e} only those  $L$'s that are not related to each other by some $M$. So we want to identify any elements $L_1$ and $L_2$ for which $\exists M \in Aut$ such that $L_1^{-1}= L_2^{-1} M$. Such an identification, rewritten as $L\sim M^{-1} L$, defines the equivalence class $[L]$ of $Aut$ in $GL(3,\mathbb{R})$ ($M$ and $M^{-1}$ take the same form); the set of these classes is precisely the coset $Aut\backslash GL(3,\mathbb{R})$, as expected. To determine this coset, we need to find representatives of each equivalence class. Consider the following generic matrix $G\in GL(3,\mathbb{R})$ and pick the $M$ with the corresponding components, then
\eq{
G=\left(
\begin{array}{ccc}
m_{11}& m_{12} & m_{13}\\
m_{21}& m_{22} & m_{23}\\
m_{31} & m_{32} &m_{33}
\end{array}\right)
\Longrightarrow
M^{-1}\cdot G= \left(
\begin{array}{ccc}
1& 0 & \frac{m_{13}m_{22}-m_{12}m_{23}}{m_{11}m_{22}-m_{12}m_{21}}\\
0& 1 & \frac{m_{11}m_{23}-m_{13}m_{21}}{m_{11}m_{22}-m_{12}m_{21}}\\
0 & 0 &\frac{\det(G)}{(m_{11}m_{22}-m_{12}m_{21})^2}
\end{array}\right)~.
}
Requiring for this $M$ that $G=M^{-1}G$ implies that $G$ must be an element of
\eq{\label{coset}
{\cal E}=\Bigg\{ L=\left(
\begin{array}{ccc}
1& 0 & a\\
0& 1 & b\\
0 & 0 &c
\end{array}\right) ~,~~~a,b\in\mathbb{R},~c\in\mathbb{R}^* \Bigg\}
~.}
${\cal E}$ turns out to be the set of representatives. Indeed, consider first $L_1 \in {\cal E}$ and $M\in Aut$, then one can easily verify that $ML_1 \in {\cal E} \Leftrightarrow M=\id$. Consider now $L_1, L_2 \in {\cal E}$ and $L_1 \neq L_2$, then $\forall M \in Aut, \ M\neq \id,\ ML_2 \notin {\cal E}$ so $ML_2 \neq L_1$. In addition, for $M=\id,\ ML_2 = L_2 \neq L_1$. We conclude that $\forall M \in Aut,\ ML_2 \neq L_1$, so $[L_1] \neq [L_2]$. Finally, one can verify that the matrices $ML$ span $GL(3,\mathbb{R})$ since the product gives generic matrices with non-zero determinant, so ${\cal E}$ is the set of representatives of the coset $Aut\backslash GL(3,\mathbb{R})$.\footnote{Note that the matrices $L$  form a group isomorphic to $\mathbb{R}^*\ltimes\mathbb{R}^2$. However this does not mean that  $Aut\backslash GL(3,\mathbb{R})$ has the structure of a group: indeed $Aut$ is not a normal subgroup of $GL(3,\mathbb{R})$.} As a side remark, for the torus one gets ${\cal E}=\lbrace \id_3 \rbrace$.

We now want to act with this set of transformations in \eqref{e}. The inverse matrices in \eqref{coset} being of the same form, we finally obtain from (\ref{heism}) and (\ref{e}) the most general form of the metric
\eq{\label{expl}\d s^2=\big(e^1+a e^3\big)^2+\big(e^2+b e^3\big)^2+c^2\big(e^3\big)^2~,~~~a,b\in\mathbb{R},~c\in\mathbb{R}^*~.}
We deduce that $\sqrt{g}= r^1 r^2 r^3 |c|$. Hence the volume is given by
\eq{\label{vol}
V=\int\d^3 x\sqrt{g}= r^1 r^2 r^3 |c|
~,}
where we used that $x^m\in[0,1]$.

Let us finally discuss the geometric meaning of the metric parameters $a$, $b$, $c$ present in \eqref{expl} besides the three radii $r^m$ and the twist $N$ of \eqref{a}. First note that $c$ is not a true parameter and can be set to $c=1$ without loss of generality. Indeed $c$ can be absorbed in the remaining parameters by the rescaling
\eq{a\rightarrow a |c|~;~~~b\rightarrow b |c|~;~~~r^3\rightarrow \frac{r^3}{|c|}~;~~~\f\rightarrow \frac{\f}{|c|}~,}
as can easily be seen from \eqref{a} and \eqref{expl}. For notation convenience we shall keep $c$ explicit in the following though, but let us set $c=1$ for the remainder of this section. The remaining parameters $a$, $b$ can be thought of as analogues of the complex structure or ``angle'' parameters of an untwisted torus. To show this, let us restrict the $S^1$ fibration over the circle of the $T^2$ base parameterized by $x^1$ (by fixing $x^2$ to a constant value) and set $b=0$, $N=0$. The resulting space is an untwisted two-torus with metric
\eq{\d s^2= (r^1)^2\left( \left(\d x^1+a\frac{r^3}{r^1}\d x^3\right)^2+ \left(\frac{r^3}{r^1}\d x^3\right)^2 \right)~.}
On the other hand, the metric of a two-torus with area $v$ and complex structure $\tau=\tau_1+i\tau_2$ is given by
\eq{\d s^2= \frac{v}{\tau_2}\left( (\d x^1+\tau_1\d x^3)^2+ (\tau_2\d x^3)^2 \right)~.}
Comparing the two metrics above we obtain: $v=r^1r^3$, $\tau_1=a r^3/r^1$, $\tau_2=r^3/r^1$. The two sets of parameters ($r^1$, $r^3$, $a$) and ($v$, $\tau$) thus provide equivalent  parameterizations of the torus, and in particular $a$ is given by the ratio $\tau_1/\tau_2$. A completely analogous interpretation can be given for $b$.

\section{Kaluza-Klein spectrum for a scalar field} \label{sec:KK}

Given the explicit form
of the metric \eqref{expl}, we will calculate the Laplacian of a scalar $\Phi$
\eq{\label{Lap}  \nabla^2\Phi=\frac{1}{\sqrt{g}}
\partial_m\left(\sqrt{g}g^{mn}\partial_n\Phi\right)
~.}
The determinant of the metric being constant, it drops out here.

\subsection{Simplest case: $a=b=0$}
\label{sec:simple}

As a warm-up we start by considering the special case $a=b=0$, and we set all other parameters to unity: $c=1$, unit radii $r^1=r^2=r^3=1$ and $\f=1$. Then the metric \eqref{expl} becomes
\eq{\label{expl2}\d s^2=\big(\d x^1)^2+\big(\d x^2)^2+\big(\d x^3+x^1\d x^2\big)^2~,}
which gives the Laplacian
\eq{\label{lapl}\nabla^2u=\left(\partial_1^2 + (\partial_2 - x^1 \partial_3)^2 + \partial_3^2 \right)u~.}
In this paper we limit ourselves to studying a scalar field, thus the wave-functions are simply eigenmodes of the above Laplacian.

We would like to expand $u$ on the space of functions invariant under \eqref{heisids}. For functions that do not depend on the coordinate $x^3$, the Laplacian can easily be diagonalised
\eq{\label{laplul0}
\big(\nabla^2+\mu^2_{p,q}\big)\tilde{v}_{p,q}=0
~,}
where the Klein-Gordon masses $\mu_{p,q}$ are given by
\eq{\label{kgm0}
\mu^2_{p,q}=4\pi^2( p^2+q^2)
~,}
and we defined
\eq{\label{uult0}
\tilde{v}_{p,q}(x^1,x^2)= e^{2\pi p \i x^1}e^{2\pi q \i x^2}
~;~~~p,q\in\mathbb{Z}~,
}
which are manifestly invariant under (\ref{heisids}).

More generally a basis of invariant functions $u_{k,l}$ can be expressed in terms of Weil-Brezin-Zak transforms \cite{thang}
\eq{\label{wbz}
u_{k,l}(x^1,x^2,x^3)=e^{2\pi k \i(x^3+x^1x^2)}
e^{2\pi l \i x^1}\sum_{m\in\mathbb{Z}}e^{2\pi k m\i x^1}f(x^2+m)
~;~~~k,l\in\mathbb{Z}~.}
It can be checked that these functions $u_{k,l}$ are indeed invariant under \eqref{heisids} for any  $f(x)$, so they are well-defined on our manifold. Plugging (\ref{wbz}) into the Laplacian (\ref{lapl}) we obtain
\eq{\label{laplred}
\nabla^2u_{k,l}= e^{2\pi k \i(x^3+x^1x^2)} e^{2\pi l \i x^1} \sum_{m\in\mathbb{Z}} e^{2\pi k m\i x^1} \Big\{ \partial^2_2  -4\pi^2 \Big( k^2 +  \left(k (x^2+ m) + l \right)^2 \Big) \Big\} f(x^2 + m)
~,}
where we consider $k\neq0$ to maintain an $x^3$ dependence. Setting $z_m=x^2+m+l/k$ and $g(z_m)=f(x^2+m)$ this becomes
\eq{\label{laplred2}
\nabla^2u_{k,l}= e^{2\pi k \i(x^3+x^1x^2)} e^{2\pi l \i x^1} \sum_{m\in\mathbb{Z}} e^{2\pi k m\i x^1}\Big\{ \partial^2_{z_m}  -(2\pi k)^2 ( z_m^2 + 1) \Big\} g(z_m)
~.}
Let us now recall the definition of the normalised Hermite functions
\eq{\label{sh}
\Phi_n(z)=e^{-\frac12 z^2}H_n(z)
~;~~~n\in\mathbb{N}~,}
where $H_n$ are the Hermite polynomials.\footnote{Recall that the Hermite polynomials are defined as $H_n(y)=(-1)^n e^{y^2} \del^n_y e^{-y^2}$ and satisfy the differential equation $\del_y^2 H_n - 2 y \del_y H_n + 2n H_n=0$.} Following \cite{thang}, for $\lambda\in\mathbb{R}^*$ we define
\eq{\label{md}
\Phi^{\lambda}_n(z)=|\lambda|^{\frac14}\Phi_n(|\lambda|^{\frac12}z)
~,}
which obeys the differential equation
\eq{\label{feq}
( \partial^2_{z}  -\lambda^2 z^2 )\Phi^{\lambda}_n(z) =
-(2n+1)|\lambda|
\Phi^{\lambda}_n(z)
~.}
Hence by setting $g(z_m)=\Phi^{2\pi k}_n(z_m)$ in (\ref{laplred2}) we obtain:
\eq{\label{laplul}
\big(\nabla^2+M^2_{k,l,n}\big)\tilde{u}_{k,l,n}=0
~,}
where the Klein-Gordon masses $M_{k,l,n}$ are given by
\eq{\label{kgm}
M^2_{k,l,n}=(2\pi k)^2\left(1+\frac{2n+1}{2\pi |k|}\right)
~,}
and we defined
\eq{\spl{\label{uult}
\tilde{u}_{k,l,n}(x^1,x^2,x^3)=e^{2\pi k \i(x^3+x^1x^2)}
e^{2\pi l \i x^1}\sum_{m\in\mathbb{Z}}& e^{2\pi k m\i x^1} \Phi^{2\pi k}_n(x^2+m+\frac{l}{k})
~;\\
&\ \ l=0,\dots, |k|-1~,~k\in\mathbb{Z}^*~,~n\in\mathbb{N}.
}}
There is a mass degeneracy since \eqref{kgm} is independent of $l$. However the wave-functions \eqref{uult} are parameterised by a finite number of inequivalent $l$'s, $0 \leq l \leq |k|-1$, i.e. the level of degeneracy is $|k|$.
We have restricted $l$ to the range of values indicated above due to the fact that $l$ is only defined modulo $k$:\footnote{We remark that $l$ being bounded by the frequency $k$ along $x^3$ is reminiscent of the integer condition \eqref{eq:n1n3} on the winding modes, obtained via consistency of the compact space.} this is a consequence of the identity
\eq{\label{lmod}
\tilde{u}_{k,l+pk,n}(x^1,x^2,x^3)=\tilde{u}_{k,l,n}(x^1,x^2,x^3)
~,~~~\forall p\in\mathbb{Z}~,}
which readily follows from the definition of $\tilde{u}_{k,l,n}$.  Finally, we remark that the only zero mode (with vanishing mass) is given by the wave-function $\tilde{v}_{0,0}$, thus it belongs to the modes on the torus base (see a related point around \eqref{0mode}).\\

In order to have a physical spectrum, we now reintroduce the dimensional parameters mentioned in Section \ref{sec:explicit}, namely $r^m, \f$, and we also want to include $c$. We obtain
\eq{\label{masses}\spl{
M^2_{k,l,n} &= k^2 \, \left(\frac{2\pi }{r^3 c}\right)^2 +(2n+1) |k|\, \frac{2\pi |\f|}{r^3 } \ , \\
\mu_{p,q}^2 &= p^2 \, \left(\frac{2\pi}{r^1}\right)^2 + q^2\, \left(\frac{2\pi}{r^2}\right)^2 \ ,
}}
while the orthonormal modes are given by
\eq{\label{uv}\spl{
u_{k,l,n}(x^1,x^2,x^3)&= \sqrt{\frac{r^2}{|N|V}}\frac{1}{\sqrt{2^n n! \sqrt{\pi}}}\,  e^{2\pi \i k (x^3+N\, x^1x^2)} e^{2\pi \i l x^1}\sum_{m\in\mathbb{Z}} e^{2\pi \i k m x^1} \Phi^{\lambda}_n(w_m) \ ,\\
v_{p,q}(x^1,x^2)&= \frac{1}{\sqrt{V}}\, e^{2\pi \i p x^1 } e^{2\pi \i q x^2 } \ ,
}}
with $\lambda= k \frac{c}{|c|} \frac{2\pi \f}{r^3}$ and $w_m=r^2 \left( x^2 + \frac{m}{N} + \frac{l}{kN} \right) $. These results are compatible with those obtained in the general case, derived in the next section.

\subsection{Most general case}
\label{sec:specgen}

The vielbeins $e^a{}_m$ giving the one-forms \eqref{a} can be written in terms of the matrix $e$
\eq{\label{33}e=\left(\begin{array}{ccc} r^1 & & \\ & r^2 & \\  & r^3 N x^1 & r^3 \end{array}\right) \ ,\ e^a=e^a{}_m \d x^m \ ,}
and we denote $(e^{-1})^m{}_a\equiv e^m{}_a$. The general metric \eqref{expl} can be written as a matrix $g=(Le)^T \delta Le$ for $L\in {\cal E}$ \eqref{coset} with $\delta=\id_3$. It is easy to verify the value of $\sqrt{g}$ given below \eqref{expl}. The inverse metric is then simply $g^{-1}=(Le)^{-1} \delta^{-1} (Le)^{-T}$. The vector $e^{-T} \del$, of component $(e^{-T})_a{}^m \del_m=e^m{}_a \del_m$, is the dual vector or co--frame to the above one-forms; analogously, $(Le)^{-T}\del=L^{-T}e^{-T}\del$ is the co--frame to $E^a$ in \eqref{heism} (the $L$ here is the inverse of the one in \eqref{e}). Using \eqref{a}, those vectors are given by
\eq{(Le)^{-T}\del = \left(\begin{array}{l} \frac{\del_1}{r^1} \\  \frac{\del_2}{r^2} - \f\, r^1 x^1 \frac{\del_3}{r^3} \\ \frac{1}{c}\left(-a\frac{\del_1}{r^1} - b \frac{\del_2}{r^2} + (1+\f\, br^1 x^1)\frac{\del_3}{r^3} \right) \end{array}\right) \ .\label{vectors}}
Since the determinant of the metric is constant, and because of the easily checked property $\del_m e^m{}_a=0$ (without sum), the Laplacian \eqref{Lap} is simply given by the square (with $\delta^{-1}$) of these vectors
\eq{\nabla^2 \Phi= \left((Le)^{-T}\del\right)^2 \Phi~, \label{Laplacianvectors}}
that results here in
\eq{\nabla^2 = \left( \frac{\del_1}{r^1} \right)^2 + \left( \frac{\del_2}{r^2} -\f\, r^1 x^1 \frac{\del_3}{r^3} \right)^2 + \frac{1}{c^2} \left( -a\frac{\del_1}{r^1} - b \frac{\del_2}{r^2} + (1+\f\, br^1 x^1)\frac{\del_3}{r^3} \right)^2 \ .}
Introducing $X^m=r^m x^m$, the Laplacian reads
\eq{\label{laplcap}\nabla^2 = \left(\del_{X^1} \right)^2 + \left( \del_{X^2} - \f\, X^1 \del_{X^3} \right)^2 + \frac{1}{c^2} \left(\del_{X^3} -a \del_{X^1} - b \left( \del_{X^2} - \f\, X^1 \del_{X^3} \right) \right)^2 \ .}
In terms of the $X^m$'s the discrete identifications (\ref{heisids}) take the form
\eq{\label{heisidscap}
X^1\sim X^1+n^1r^1~;~~~X^2\sim X^2+n^2r^2~;~~~X^3\sim X^3+n^3r^3- \f\, n^1r^1 X^2~,~~~n^1,n^2,n^3\in\{0,1\}~.
}
The following set of functions is invariant under these identifications, thanks to \eqref{quant},
\eq{\label{wbzcap}
U_{k,l}(X^1,X^2,X^3)=e^{2\pi K \i(X^3+\f\, X^1X^2)}
e^{2\pi L \i X^1}\sum_{m\in\mathbb{Z}}e^{2\pi K M\i X^1}F\left(X^2+\frac{M}{\f}\right)
~,}
where
\eq{\label{la}K=\frac{k}{r^3}~,~~L=\frac{l}{r^1}
~,~~ M=\frac{r^3}{r^1}m~;~~~m,k,l\in\mathbb{Z}~.}
Now we follow a similar procedure as in the simple case before. We consider $K\neq 0$. Plugging (\ref{wbzcap}) into the Laplacian (\ref{laplcap}), setting $z_m=X^2+M/\f+L/(K\f)$ and $G(z_m)=F\left(X^2+\frac{M}{\f}\right)$ we obtain
\eq{\spl{\label{laplred2cap}
\nabla^2U_{k,l}= & e^{2\pi K \i(X^3+\f\, X^1X^2)} e^{2\pi L \i X^1}\sum_{m\in\mathbb{Z}} e^{2\pi K M\i X^1}\\ & \qquad \times\Big\{ \partial^2_{z_m}  -(2\pi K\f)^2 z_m^2
+\frac{1}{c^2}\left(
2\pi K \i (1 - \f a z_m) -b\partial_{z_m}
\right)^2
\Big\} G(z_m)
~.}}
The above can also be rewritten as
\eq{\spl{\label{laplred3cap}
\nabla^2U_{k,l}= & e^{2\pi K \i(X^3+\f\, X^1X^2)} e^{2\pi L \i X^1}\sum_{m\in\mathbb{Z}} e^{2\pi K M\i X^1}\exp\left[-\frac{\i\pi K b}{b^2+c^2}\ z_m (\f a z_m- 2) \right]\\
&\quad \times
\Big\{ \partial^2_{z_m}  -
\frac{4\pi^2K^2c^2}{(b^2+c^2)^2}\left[
(1-\f a z_m)^2
+\f^2 (b^2+c^2)z_m^2
\right]
\Big\} H(z_m)
~,}}
where we have defined
\eq{H(z)=\frac{b^2+c^2}{c^2}\exp\left[\frac{\i\pi K b}{b^2+c^2}\ z (\f a z- 2)\right]G(z)~
.}
With a further change of variable
\eq{w_m=z_m-\frac{a}{\f(a^2+b^2+c^2)}~,}
equation \eqref{laplred3cap} becomes (for convenience we do not change notation in the first row: $z_m$ should be thought of as a function of $w_m$)
\eq{\spl{\label{laplred4cap}
\nabla^2U_{k,l}= & e^{2\pi K \i(X^3+\f\, X^1X^2)} e^{2\pi L \i X^1}\sum_{m\in\mathbb{Z}} e^{2\pi K M\i X^1}
\exp\left[-\frac{\i\pi K b}{b^2+c^2}\ z_m (\f a z_m- 2) \right]
\\
& \qquad \times\Big\{ \partial^2_{w_m}  -
\frac{4\pi^2K^2c^2}{(b^2+c^2)^2}\left[
\frac{b^2+c^2}{a^2+b^2+c^2}
+\f^2 (a^2+b^2+c^2)w_m^2
\right]
\Big\} T(w_m)
~,}}
where $T(w_m)=H(z_m)$. Finally substituting in (\ref{laplred4cap}) $T(w_m)=\Phi^{\lambda}_n(w_m)$,
with
\eq{\label{lambda}
\lambda=\frac{2\pi K c}{b^2+c^2}(a^2+b^2+c^2)^{\frac12} \f
~,}
taking (\ref{feq}) into account, we obtain
\eq{\label{laplulcap}
\big(\nabla^2+ M^2_{k,l,n}\big)U_{k,l,n}=0
~,}
where the Klein-Gordon masses $M_{k,l,n}$ are given by
\eq{\label{kgmcap}
M^2_{k,l,n}=\frac{4\pi^2 k^2}{(r^3)^2(a^2+b^2+c^2)}\left[1+\frac{(2n+1)r^3|\f|}{2\pi |k c|}(a^2+b^2+c^2)^{\frac32}\right]
~,}
and we defined
\eq{\spl{\label{uultcap}
U_{k,l,n}(x^1,x^2,x^3)&= \sqrt{\frac{r^2}{|N|V}}\frac{1}{\sqrt{2^n n! \sqrt{\pi}}}\,  e^{2\pi K \i(X^3+\f\, X^1X^2)} e^{2\pi L \i X^1}\sum_{m\in\mathbb{Z}} e^{2\pi K M\i X^1}
\\
\times&\exp\left[-\frac{\i\pi K b}{b^2+c^2}\ z_m (\f a z_m- 2) \right] \Phi^{\lambda}_n(w_m) ~;
~~~k\in\mathbb{Z}^*~,n\in\mathbb{N}.
}}
Note that the $U_{k,l,n}$ differ from the $U_{k,l}$ by a factor $\frac{b^2 +c^2}{c^2} \sqrt{\frac{r^2}{|N|V}}\frac{1}{\sqrt{2^n n! \sqrt{\pi}}}$, $V$ being the volume \eqref{vol}. This has no influence on the mass, but allows the $U_{k,l,n}$ to be orthonormal as verified in Appendix \ref{ap:ortho}. We rewrite the above as (with $\lambda$ given in \eqref{lambda})\footnote{As a side remark, we indicate that the product of exponentials can be rewritten as
\eq{\exp\left[\frac{2\i\pi K}{b^2+c^2} c^2 (X^3 +\f\, X^1 z_m)\right] \exp\left[\frac{2\i\pi K}{b^2+c^2} b\left( b (X^3 +\f\, X^1 z_m) -\frac{\f a}{2} z_m^2 +z_m \right) \right]\ ,\nn}
where each of those two exponentials gets acted on non-trivially by only one of the three terms in the Laplacian \eqref{laplcap}. This might be of interest in  generalizing the solution to other nilmanifolds.}
\eq{\spl{\label{uultcap2}
U_{k,l,n}(x^1,x^2,x^3)&=  \sqrt{\frac{r^2}{|N|V}}\frac{1}{\sqrt{2^n n! \sqrt{\pi}}}\, e^{2\pi K \i(X^3+\f\, X^1X^2)} e^{2\pi L \i X^1}\sum_{m\in\mathbb{Z}} e^{2\pi K M\i X^1} \\
&  \times \exp\left[-\frac{\i\pi K b}{b^2+c^2} \bigg(X^2 + \frac{M}{\f} + \frac{L}{K\f} \bigg) \bigg(\f a \bigg(X^2 + \frac{M}{\f} + \frac{L}{K\f} \bigg) - 2 \bigg) \right] \\
&  \times \Phi^{\lambda}_n\bigg(X^2 +\frac{M}{\f} + \frac{L}{K\f} - \frac{a}{\f(a^2+b^2+c^2)}\bigg)
~;\\
& \qquad \qquad \qquad \qquad \qquad \qquad l=0,\dots, |k|-1~,~k\in\mathbb{Z}^*~,n\in\mathbb{N} \ .
}}
As in Section \ref{sec:simple} the range of $l$ is finite, leading to a finite degeneracy in the masses. This can be seen from the invariance of (\ref{uultcap2}) under $L\rightarrow L+KP$, where $P=\frac{r^3}{r^1}p$ with $p\in\mathbb{Z}$: taking (\ref{la}) into account it then follows that (\ref{uultcap2}) is invariant under $l\rightarrow l+kp$. It is also straightforward to recover the results of Section \ref{sec:simple} by setting the parameters to the appropriate values.

The masses obtained in \eqref{kgmcap} only depend on the radius of the fiber, $r^3$ (or $r^3 c$). This is a way to understand the modes just discussed as coming from the fiber. As in the simple case, there should be other modes coming from the base, independent of the fiber coordinate. The most general decomposition of such modes, invariant under the identifications \eqref{heisidscap}, can be given in terms of a Fourier basis, i.e. in orthonormal form
\eq{\label{52}V_{p,q}(x^1,x^2)= \frac{1}{\sqrt{V}}\, e^{2\pi \i P X^1 } e^{2\pi \i Q X^2 } \ , \quad P=\frac{p}{r^1},\ Q=\frac{q}{r^2},\ p,q \in \mathbb{Z}\ .}
One then gets
\eq{\label{53} (\nabla^2 + \mu_{p,q}^2)  V_{p,q}=0 \ , \quad \mu_{p,q}^2=4\pi^2\left(\frac{p^2}{(r^1)^2} + \frac{q^2}{(r^2)^2} + \frac{1}{c^2} \left(a \frac{p}{r^1} + b \frac{q}{r^2} \right)^2  \right) \ .}
As expected, these masses only depend on the base radii. The modes $V_{p,q}$ and $U_{k,l,n}$ form the complete set of eigenmodes of the Laplacian on $\mcal_3$, as verified in Appendix \ref{ap:complete}.

\section{Numerical study of the spectrum} \label{sec:num}

Having an analytical solution for the KK spectrum on a non-trivial manifold is an interesting opportunity to check the validity of numerical methods that could be used in situations where a complete derivation is out of reach, or as a cross-check on the completeness of the set of modes analytically obtained. In this section we will check the validity of our implementation of a numerical method that determines the eigenvalues of the Laplacian. We will use the results derived in Section \ref{sec:simple} in the simplest case where $a=b=0$, $c=1$, $r^{m=1,2,3}=1$ and $\f=1$.

\subsection*{Principle of the analysis}

Several methods have been proposed to find the spectrum of Laplace operators numerically, in particular on hyperbolic manifolds \cite{Cornish,Cornish2,Inoue} where no analytic results are known. We found the algorithm by Cornish and Turok \cite{Cornish} to be the most straightforward to adapt to nilmanifolds since it does not require any data apart from the Laplacian operator \eqref{lapl} and a definition of the fundamental domain of the compactified manifold \eqref{heisids}. Let us first review the principle of the algorithm:

\begin{itemize}
\item define a lattice representing the geometry of a choice of fundamental domain;
\item define an approximate Laplacian operator $\Delta$ on this lattice, using appropriate boundary conditions to evaluate derivatives on the edge of the lattice;
\item define an initial condition $\phi(x,t=0)=\phi_0(x)$ for the field (for simplicity we always take $\del_t \phi(x,t=0) = 0$);
\item generate the time evolution of this field according to the equation $\del_t^2 \phi = \Delta \phi$.
\end{itemize}
The rationale is that the wave equation has a set of solutions given by $\phi(x,t)=\psi_q(x) e^{i q t}$ where $\psi_q$ is an eigenfunction of $\Delta$ with eigenvalue $-q^2$ so that if $\phi_0(x)=\sum \alpha_q \psi_q(x)$, the solution can be expressed as $\phi(x,t) = \sum \alpha_q \psi_q(x) e^{i q t}$. Provided the spectrum is discrete (the case for a compact space), the field has a Fourier transform in the time variable composed of Dirac peaks whose positions are given by the spectrum of the Laplace operator. A second step in the algorithm is therefore to post-process the solution $\phi(x,t)$ as follows:

\begin{itemize}
\item perform a point-wise Fourier transform to get $\hat \phi(x,\omega) = \int \d t \phi(x,t) e^{-i\omega t}$;
\item extract the power spectrum $C(\omega) = \int_{{\cal M}_3} \d^3 x \sqrt{g}  \left|\hat \phi (x,\omega)  \right|$;
\item extract the peaks to obtain a numerical spectrum.
\end{itemize}

\subsection*{Details of the implementation}

In the simple case studied, the fundamental domain is a cube of physical dimension $1\times 1 \times 1$ with boundary conditions straightforwardly derived from \eqref{heisids}:

\begin{itemize}
\item $x^2 \in \left[0,1\right[$ and $(x^1,0,x^3)\sim (x^1,1,x^3)$,
\item $x^3 \in \left[0,1\right[$ and $(x^1,x^2,0)\sim (x^1,x^2,1)$,
\item $x^1 \in \left[0,1\right[$ and $(0,x^2,x^3)\sim (1,x^2,x^3-x^2)$.
\end{itemize}
A cubic lattice on this fundamental domain is therefore perfectly suited for our purpose: it is stable under the boundary conditions for any value of the spacing $\delta x$.
We performed the time evolution using the leapfrog method, which relies on the ${\cal O}(\delta t^2)$ approximation of the time derivative in the wave equation:
\begin{equation}
  \label{eq:leapfrog}
  \phi(x,t+\delta t) \simeq 2 \phi(x,t) -\phi(x,t-\delta t) + \delta t \Delta \phi(x,t)~,
\end{equation}
where the Laplace operator is computed using the ${\cal O}(\delta x^2)$ centred finite difference definition of the second order derivatives in \eqref{lapl}. Whenever a derivative requires the value of the field outside the fundamental domain, the boundary conditions are used to provide a value from within the fundamental domain.
At this order of precision in time and lattice spacing, the leapfrog method in flat space is proved to be stable under the condition that $\delta t<\frac{1}{\sqrt{3}}\delta x$ and, while the proof does not carry on to our situation, it seems indeed that, at small enough time spacings, stability is ensured to very long simulation times on nilmanifolds as well. In practice, we used a lattice spacing $\delta x = 0.01$ and a time spacing $\delta t = 0.005$, which allowed simulations to be run to $t = 100$.

The Fourier transforms were calculated using the Gnu Scientific Library \cite{Gough} implementation of the discrete Fourier transform, which takes as input a sampled function $\{ f(k\Delta t)\}_{0\leq k\leq n}$ at a rate $\Delta t$ over a period $T=n\Delta t$ and returns a discrete spectrum $\{\hat f (l \Delta \omega)\}_{0\leq l\leq n/2}$, with resolution $\Delta \omega = 2\pi /T$ and range up to a maximum pulsation of $\omega = 2\pi /{\Delta t}$. The long evolution time we used in our simulations therefore resulted in a good resolution in the spectra we computed, ensuring a sufficient separation between the different peaks.

\subsection*{Results}

\begin{figure}[!h]
  \centering
\includegraphics[width=0.5\textwidth]{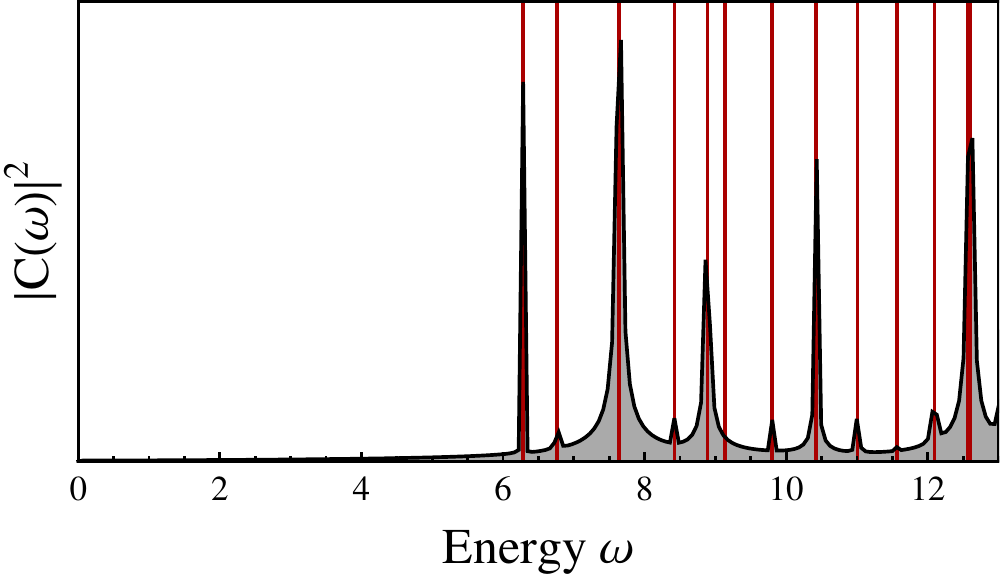}\includegraphics[width=0.5\textwidth]{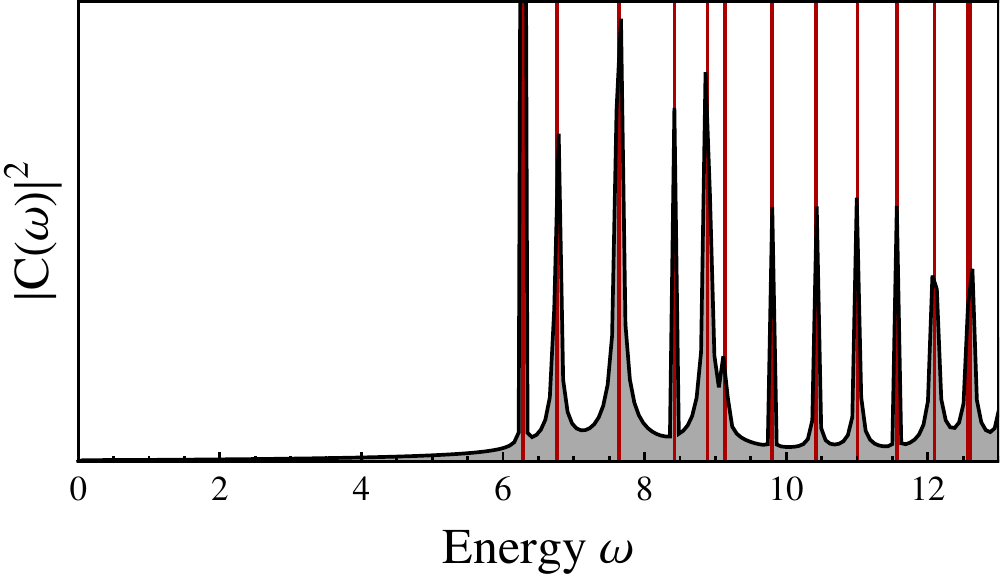}\\
\hspace{3em}(a)\hspace{0.48\textwidth}(b)
  \caption{Fourier spectra found for two initial conditions: (a) wide gaussian, (b) narrow gaussian. The black curve is the numerical spectrum and the vertical red lines indicate the position of the eigenvalues derived analytically in Section~\ref{sec:KK}. The vertical axes have arbitrary units. As expected, the narrow Gaussian excites a larger number of modes than the wide Gaussian.}
  \label{specs}
\end{figure}

To obtain a spectrum, we used initial states $\phi_0(x)$ generated using randomly centred gaussians within the central region of the fundamental domain. The standard deviation of these functions gave us a handle to probe different length scales while ensuring that the boundary conditions were verified by having $\phi_0(x)\simeq 0$ on all the edges of the fundamental domain. As can be seen in Figure \ref{specs}, we find very good agreement with the low-energy spectrum derived analytically in \eqref{kgm0} and \eqref{kgm}.

\begin{figure}[!h]
  \centering
\includegraphics[width=0.5\textwidth]{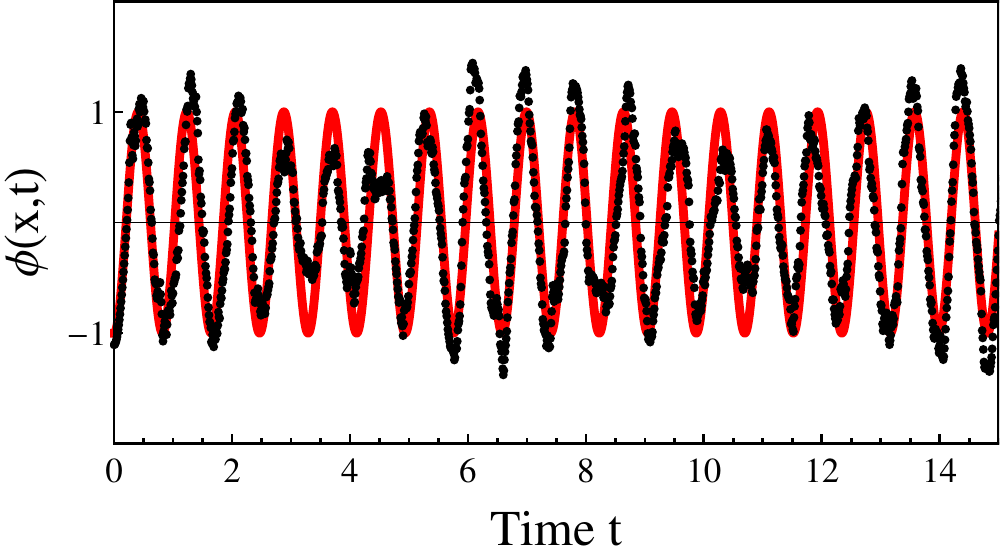}\includegraphics[width=0.5\textwidth]{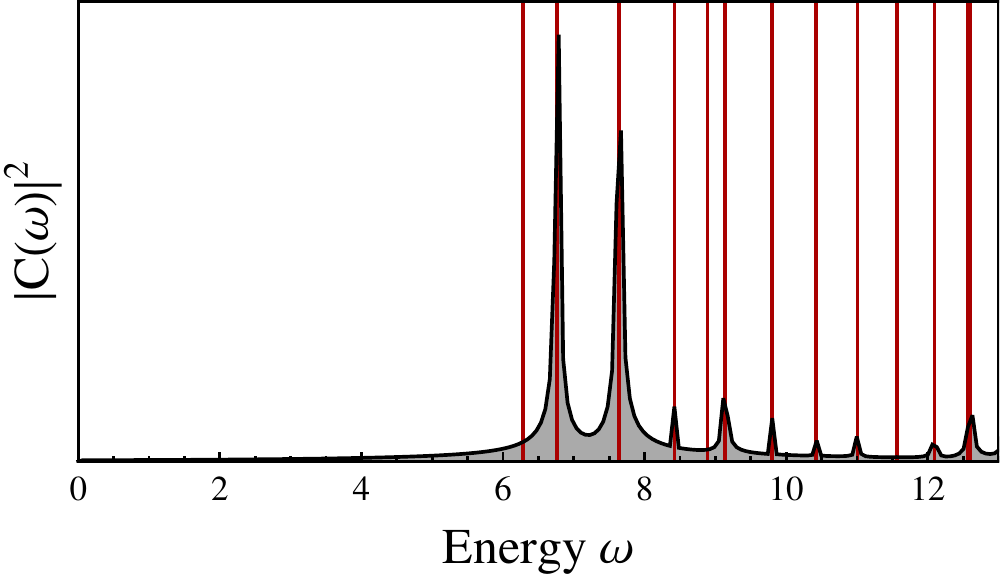}\\
\hspace{3em}(a)\hspace{0.48\textwidth}(b)
  \caption{Analysis of the $\tilde u_{1,0,1}+\tilde u_{-1,0,1}$ eigenmode initial state. (a) The short-time evolution of the wave-function at a given point $x_0$ exhibits an approximate sinusoidal behaviour with the appropriate pulsation $\omega_e=7.63$ ($\phi(x_0,t)$ in dotted black, $-\cos(\omega_e t)$ in continuous red). (b) The Fourier spectrum of the field over an evolution time of 100. A significant leaking has happened to the lower mode $\tilde u_{1,0,0} + \tilde u_{-1,0,0}$ due to the long evolution time needed to produce a well-resolved spectrum.}
  \label{eigenplot}
\end{figure}

Another check of the validity of the implementation is to use the eigenfunctions of the Laplacian as an initial state. As seen in their expression \eqref{uult}, the combination $\tilde u_{k,0,n}+\tilde u_{-k,0,n}$ is real and contains a series with a fast convergence due to the exponential in the definition of the Hermite functions \eqref{sh}. A truncation of this series gives therefore a good approximation of its values and we could check that such an initial state does show the expected harmonic oscillation as depicted in Figure \ref{eigenplot} (a). At the level of discretisation we used for our simulations, however, this pure frequency behaviour is not stable for long evolution times and significant leaking to neighbouring modes happens, leading to the double peak feature seen in Figure \ref{eigenplot} (b).

In light of these results, we remark that the algorithm we used to obtain the low-energy scalar KK spectrum on the simplest nilmanifold works precisely enough to be used as a predictive tool in contexts where an analytical solution is difficult or impossible to obtain. In particular, it could be applied to find the spectrum of gauge fields and fermions in particle physics models with nilmanifold extra-dimensions without localised fields.

\section{Isometries, orbifolding, and Dark Matter} \label{sec:iso}

We first study in this section two discrete isometries of our Heisenberg nilmanifold $\mcal_3$. Considering the resulting orbifolds, we discuss their fixed points. Using these results and the above KK spectrum, we propose a simple model for Dark Matter.

\subsection{Orbifolding and fixed points}

We look for discrete isometries of the nilmanifold, such that an
orbifolding can be performed. Explicitly we consider transformations of the form:
\eq{\label{trns}\left(
\begin{array}{c}
x^1\\
x^2\\
x^3
\end{array}\right) \rightarrow \left(
\begin{array}{c}
x^{\prime1}\\
x^{\prime2}\\
x^{\prime3}
\end{array}\right)
=J\cdot \left(
\begin{array}{c}
x^1\\
x^2\\
x^3
\end{array}\right)
~,}
where $J\in GL(3,\mathbb{R})$ is a constant matrix (more generally, $J$ corresponds to $\del x'/\del x$). The isometry condition then takes the form (in matrix notation):
\eq{\label{iso}
g(x)=J^{T}\cdot g(x')\cdot J
~.}
The metric can be written as
\eq{g(x)=(L\cdot e(x))^T\cdot L\cdot e(x)~,}
where the matrices $L$, $e$ were defined in \eqref{coset} and \eqref{33} respectively.

\subsubsection*{Transformation $T$}

Rather than examining the most general case, we first focus on discrete actions on the $T^2$ base of the fibration, such that the fiber coordinate $x^3$ remains invariant.
It can then be seen that there is only one nontrivial solution to \eqref{iso}, corresponding to the transformation
\eq{\label{trns2}
T:~~ x^{1,2} \rightarrow
-x^{1,2}
~;~~~x^3\rightarrow x^{3}~,}
and we must set $a=b=0$; all other parameters of the metric can be arbitrary.
The resulting orbifolded base $T^2/\mathbb{Z}_2$ is a square, topologically a disc: it consists of the interval parameterised by $x^1\in[0,\frac12]$ tensored with the interval parameterised by $x^2\in[0,\frac12]$. There are four fixed (fiber) circles, stemming from the points on the base: $(x^1,x^2)=(0,0),(\frac12,0),(0,\frac12),(\frac12,\frac12)$.
The scalar spectrum can be reorganised in eigenstates of the orbifold
involution (\ref{trns2}). For the torus modes, we obtain the usual spectrum of a $T^2/\mathbb{Z}_2$ orbifold: starting from \eqref{uv}, the linear combinations
\beq
\frac{1}{\sqrt{2}} (v_{l,n} \pm v_{-l, -n})\,,
\eeq
are even, odd respectively.
For the modes propagating in the fiber, it is the linear combinations
\eq{
\frac{1}{\sqrt{2}} (u_{k,l,n}\pm(-1)^n u_{k,|k|-l,n})
~,}
that are even, odd respectively. This can be seen by taking into account the definition in \eqref{uv}, and the fact that
the Hermite polynomials $H_n$ are even (odd) under parity transformations for $n$
even (odd), and we have used the identity $u_{k,-l,n}=u_{k,|k|-l,n}$.

\subsubsection*{Transformation $P$}

We would now like to perform an orbifolding of $\mathcal{M}_3$ by taking the discrete quotient with respect to the involution $P$ defined as
\eq{\label{trns3}
P:~~ x^{1} \leftrightarrow
x^{2}
~;~~~x^3\rightarrow - x^{3} - N x^1 x^2~.}
This is an isometry only for $b=-a$ in the metric, and for equal radii along the torus directions, i.e. $r^1 = r^2$. It is interesting to note that $P$ commutes with $T$ defined in \eqref{trns2}: if we define an orbifold projection on $P$, then (if $a=b=0$) $T$ acts as an exact global symmetry that may preserve some of the KK modes as stable. Here, by stable, we mean that all decays into zero modes are forbidden. As a first step, we need to define the geometry of the orbifold quotient.

Firstly let us make the definition of $P$ more precise.
For that we will make use of the open cover of $\mathcal{M}_3$ given in
Appendix \ref{sec:N} (see Figure \ref{fig3}).
$P$ exchanges $x^1$ with $x^2$, which is well-defined if $0\leq x^1, x^2<1$, since in that case $x^1\in U_+$ both before and after the transformation.
Similarly $P$ is well-defined in the case $x^1=x^2=1$.
If however $0\leq x^1<1, x^2=1$,  then $x^1\in U_+$ before the transformation but $x^1\in U_-$ after the transformation.
In that case $P$ should be understood as sending $x^1_+$ to $x^1_-=1$ and correspondingly the value of the $x^3$ coordinate before and after the
transformation should be
understood as that of $x^3_{\pm}$ respectively.
The case $0\leq x^2<1, x^1=1$ is completely analogous.
More explicitly, $P$ is defined in these two cases as
\eq{\label{trns3a}
P:~~ \left\{\begin{array}{l}
(x_+^{1},x^2,x^3_+)=(s,1,t) \longrightarrow
(x_-^{1},x^2,x^3_-)=(1,s,-t-Ns)\\
(x_-^{1},x^2,x^3_-)=(1,s,t) \longrightarrow
(x_+^{1},x^2,x^3_+)=(s,1,-t-Ns)
\end{array}
\right.
~,}
where $0\leq s<1$, $t\in \mathbb{R}$ with the understanding that $x^3\sim x^3+1$.

No potential inconsistency can arise from the definition of $P$ when $0\leq x^1, x^2<1$ since the bundle is trivial over that region of the base, being a subset of $\mathcal{B}_+\times S^1_{x^3}$. In that case  $P$ relates the fiber $S^1_{x^3}$ over a generic point $(x^1,x^2)$ with $S^1_{x^3}$ over a different  point $(x^2,x^1)$ and identifies the respective points on the base, ``folding'' the base along the diagonal in the  $x^1,x^2$-plane. Over a point on the diagonal $(x^1,x^2)=(s,s)$, $s\in[0,1[$, $P$ acts as an involution on $S^1_{x^3}$ resulting in an interval $I_s=S^1/\mathbb{Z}_2$. More precisely for each $s\in[0,1[$, $I_s$ can be parameterised by $x^3\in[-\frac12 Ns^2,-\frac12 Ns^2+\frac12]$. This can be seen from the fact that for each $(x^1,x^2)=(s,s)$, $P$ identifies the points $x^3=-\frac12 Ns^2\pm \varepsilon$ (therefore it  fixes the point $x^3=-\frac12 Ns^2$), and we have $x^3\sim x^3+1$.

\begin{figure}[tb!]
\begin{center}
\includegraphics[width=10cm]{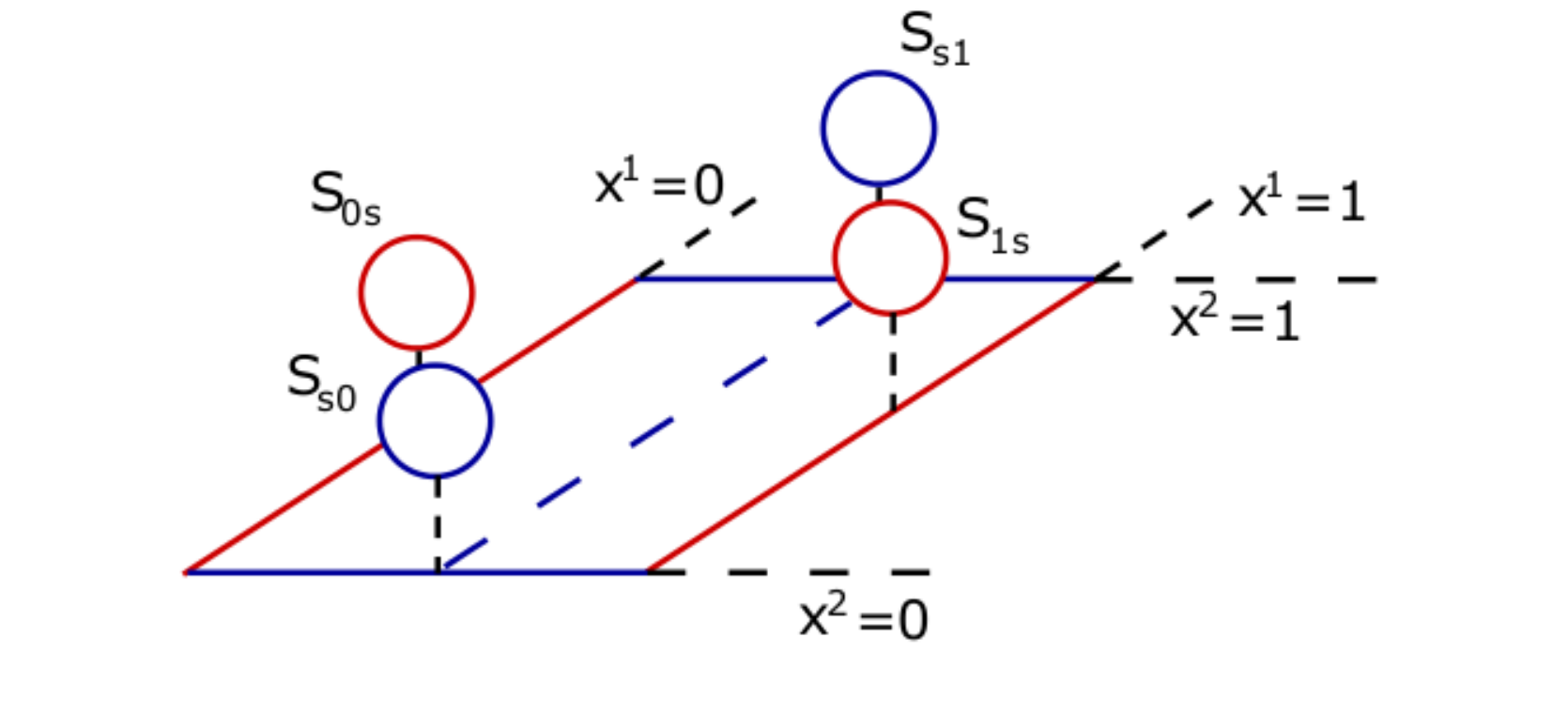}
\end{center}
\caption{The fibers over the points
$(s,1),(s,0),(1,s),(0,s)$ on the base are the circles $S_{s1}$, $S_{s0}$, $S_{1s}$, $S_{0s}$, respectively. The pairs of points $(s,1),(1,s)$ and $(s,0),(0,s)$ are mirror symmetric with respect to the diagonal joining (0,0) and (1,1). The fibers over each pair are related by the orbifolding involution $P$. Points joined by dashed lines on the base are identified.} \label{fig1}
\end{figure}

On the other hand there is a potential source of inconsistency arising from definitions (\ref{trns3a}), which can be seen as follows: for $s\in[0,1[$ define
$S_{s1}$, $S_{s0}$, $S_{1s}$, $S_{0s}$ to be the fiber over the points
$(x^1,x^2)=(s,1),(s,0),(1,s),(0,s)$ respectively, as illustrated in Figure \ref{fig1}. $S_{s1}=S_{s0}$ since the respective base points are identified and the bundle is trivial for
$0\leq x^1<1$. Moreover  $S_{s1}$ is mapped to $S_{1s}$ under (\ref{trns3a}) and $S_{s0}$ is mapped to $S_{0s}$ under (\ref{trns3}).
The consistency check is therefore that $x^3_-\in S_{1s}$ and $x^3_+\in S_{0s}$ should respect the gluing condition (\ref{gluing}) which in this case reads,
\eq{\label{gluingb}
x_{-}^3=x_{+}^3-N s
~.}
Let $x^3_{+s}$ denote the coordinate of the fiber $S_{s1}=S_{s0}$. The
mapping (\ref{trns3a}) of $S_{s1}$ to $S_{1s}$ implies: $x_{-}^3=-x_{+s}^3-N s$; similarly, the
mapping (\ref{trns3}) of $S_{s0}$ to $S_{0s}$ implies: $x_{+}^3=-x_{+s}^3$. Eliminating $x_{+s}^3$ from the previous two equations gives (\ref{gluingb}), hence the orbifolding is consistent with the fibration.

\begin{figure}[tb!]
\begin{center}
\includegraphics[width=10cm]{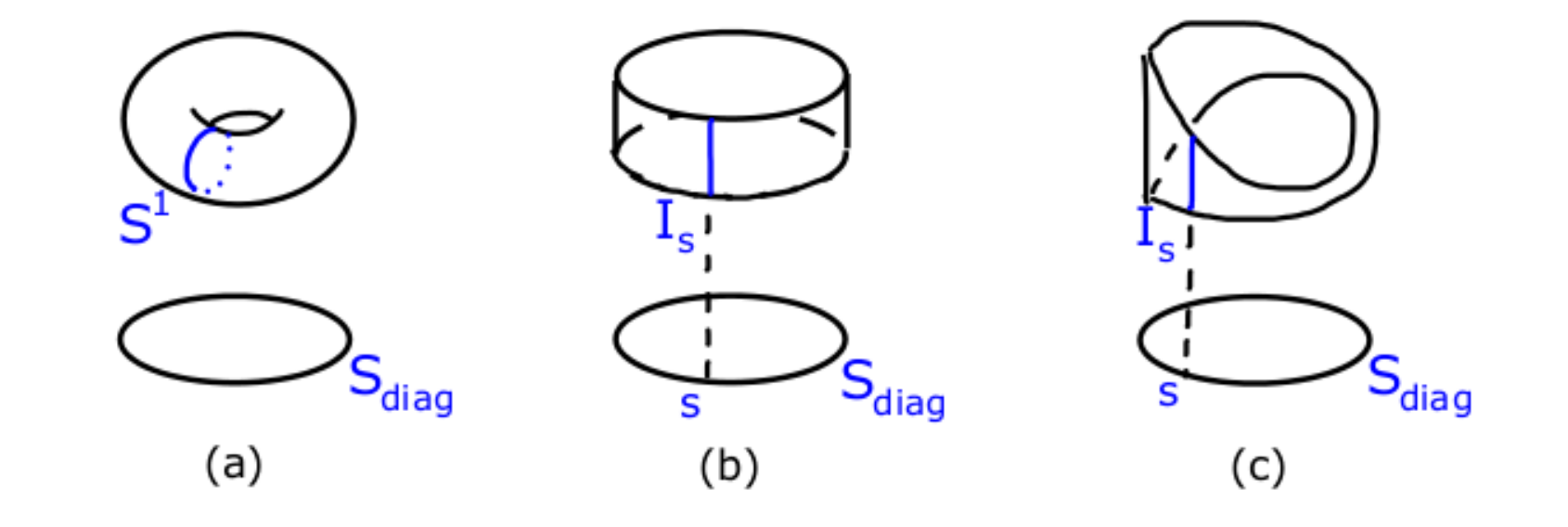}
\end{center}
\caption{(a) The $S^1_{x^3}$-fibration restricted over the diagonal  $S_{\mathrm{diag}}$ in the base is topologically a torus. (b) For $N$ even the orbifold fibration of $I_s$ over $S_{\mathrm{diag}}$ is a cylinder. The fixed-point locus is the boundary of the bundle, which consists of two circles. (c) For $N$ odd the orbifold fibration of $I_s$ over $S_{\mathrm{diag}}$ is a M\"{o}bius strip. The fixed-point locus is the boundary of the strip, which is topologically a circle.} \label{fig2}
\end{figure}

Having verified the consistency of the orbifolding action $P$, let us now determine its fixed points. As we already noted, these can only occur over the diagonal in the $x^1,x^2$-plane. Before orbifolding, let us first make some further remarks. Let $x^3_-$ be the fiber coordinate over
$(x^1_-,x^2)=(1,1)$ and let $x^3_+$ be the fiber coordinate over
$(x^1_+,x^2)=(0,1)\sim(0,0)$. In this case the gluing condition (\ref{gluing}) implies: $x_{-}^3=x_{+}^3$. It follows that, when restricted over
the diagonal in the base space, the $S^1_{x^3}$-fibration is trivial. Note that the diagonal is in fact a circle,  $S_{\mathrm{diag}}$, since the endpoints $s=0,1$ are identified.
In particular the total space of the fibration restricted over the diagonal
in the base is a trivial fibration of $S^1_{x^3}$ over $S_{\mathrm{diag}}$,
i.e. a torus (see Figure \ref{fig2}.a). This fact simplifies the analysis of the fixed-point locus when orbifolding since we can then parameterise all points in the diagonal $(x^1,x^2)=(s,s)\in S_{\mathrm{diag}}$ with a global coordinate $s\in [0,1]$.

As discussed earlier, upon orbifolding the fiber becomes $I_s$ over the diagonal base point parameterised by $s$. This fiber can be
thought of as the circle $x^3\in[0,1]$ ``folded'' along the antipodal points $x^3=-\frac12 Ns^2, -\frac12 Ns^2+\frac12$; these are the endpoints of the interval $I_s$ defined earlier. Therefore we have two lines of fixed points: $(x^1,x^2,x^3)=(s,s,-\frac12 Ns^2)$ and $(x^1,x^2,x^3)=(s,s,-\frac12 Ns^2+\frac12)$, $s\in [0,1]$. If $N$ is even, the endpoints of these lines at $s=0,1$ are  identified, so instead of two lines we have in fact two circles of fixed points. These two circles define the boundary of the $I_s$ fibration over $S_{\mathrm{diag}}$ which in this case is trivial, i.e. topologically a cylinder (see Figure \ref{fig2}.b).

If however $N$ is odd the endpoints differ by $\frac12$ in the $x^3$ coordinate.
Hence starting at one endpoint of $I_s$ over $s\in S_{\mathrm{diag}}$, after going once around $S_{\mathrm{diag}}$ we end up at the other endpoint of $I_s$. In this case the $I_s$ fibration over $S_{\mathrm{diag}}$ is therefore a M\"{o}bius strip. The fixed-point locus is the boundary of the strip, which has the topology of a (single) circle (see Figure \ref{fig2}.c).

\subsubsection*{Combining $T$ and $P$}

Finally, let us perform an orbifolding of the Heisenberg nilmanifold  by taking the discrete quotient with respect to the action generated by both involutions $T$ and $P$ defined in \eqref{trns2} and \eqref{trns3}, \eqref{trns3a} respectively.
Noting that $P$, $T$ commute, we will denote the resulting orbifold by $\widetilde{\mathcal{M}}_3=\mathcal{M}_3/\mathbb{Z}_2\times\mathbb{Z}_2$.

To determine the topology of $\widetilde{\mathcal{M}}_3$ it is easier to first mod out by
the action of $T$. As was described below (\ref{trns2}), the orbifolding by $T$ gives a three-manifold which can be thought of as an $S^1$ fibration over a square base. More precisely the base is parameterised by $(x^1,x^2)$ and it is a square with vertices $(x^1,x^2)=(0,0),(\frac12,0),(0,\frac12),(\frac12,\frac12)$. The $S^1$ fiber is parameterised by $x^3$. After modding out by $T$ there are no remaining discrete
identifications among the base points, hence the $x^1$,$x^2$ are globally well-defined coordinates and the $S^1$ fibration is topologically trivial, i.e. modding out by $T$ gives a square (topologically a disc) tensored with a circle.

Next let us mod out by $P$. Its action on the $(x^1,x^2)$-base can be viewed as a folding of the square to form the triangle with vertices $(x^1,x^2)=(0,0),(\frac12,0),(\frac12,\frac12)$, which we denote by $O$, $A$, $B$ respectively. Let us now turn to the result of the action of $P$ on the $S^1$ fiber. Above each base point $(x^1,x^2)$ in the interior of $OAB$ we have the same circle parameterised by $x^3$ as before: the action of $P$ simply relates the circle over $(x^1,x^2)\in OAB$ to a circle over $(x^2,x^1)\notin OAB$. The same is true for the sides $OA$ and $AB$, excluding the points $O$ and $B$: over each point $(x^1,x^2)$ we have the same circle parameterised by $x^3$ as before. However on the circle $S^1$ over each point $(x^1,x^2)=(s,s)\in OB$, $s\in[0,\frac12]$, the action of $P$  results in an interval $I_s=S^1/\mathbb{Z}_2$, as mentioned previously.

To summarise: the $\widetilde{\mathcal{M}}_3$ orbifold can be described as a topologically trivial fibration with base the triangle $OAB$. Over each point on the base we have a circle fiber, except over the interval $OB$ where the fiber degenerates to an interval with endpoints $J_1$, $J_2$. The fixed locus of the $\mathbb{Z}_2\times\mathbb{Z}_2$ orbifolding consists of the two intervals: $OB\times J_1$ and $OB\times J_2$.

\subsection{A simple Dark Matter model}\label{sec:DM}

As a first example, and as a proof of concept, we want to build a model where the Dark Matter candidate is provided by a scalar field, singlet under the Standard Model (SM) gauge symmetries, which propagates in the bulk of the nilmanifold. The SM fields, on the other hand, are ordinary four-dimensional (4D) fields which propagate on a 4D subspace, i.e. a point in the nilmanifold. For this construction to be consistent, therefore, we would need an orbifold that contains singular points where a 4D brane can be localised, supporting the SM fields. The orbifold examples given above do not satisfy this requirement, as the singular points, left fixed under the orbifold symmetry, form circles or intervals in the extra space and thus correspond to 5D subspaces. For the existence of a natural Dark Matter candidate, we further require that the orbifold space possesses at least another symmetry, the Dark Matter parity, under which the KK modes can be labelled. The lightest state odd under the latter will thus be our Dark Matter candidate, as it cannot decay into zero modes nor into SM fields.

In this work, we do not attempt a complete classification of the possible orbifolds, rather we look for a simple example, i.e. the orbifold defined by the involution $P$ defined in \eqref{trns3} and \eqref{trns3a} as it commutes with $T$ in \eqref{trns2} which can be identified with the Dark Matter parity. Note that the space is characterised by $a = b = 0$ (thus corresponding to the simple case in Section \ref{sec:simple}, more precisely to \eqref{uv}), and by $r^1 = r^2 = r$. While this orbifold has no fixed points but circles, the origin $(0,0,0) \sim (1,1,1)$ plays a special role, as it belongs to the fixed points of the orbifold and it is also left fixed by $T$. Thus, localising the SM on the origin is consistent with the orbifold and it does not break the Dark Matter parity. We will therefore discuss a scenario where the SM is localised there, and the singlet bulk scalar field communicated to the SM via a Higgs portal coupling (which is the only one allowed by gauge invariance).

To recapitulate, the symmetries we use to define the orbifold and Dark Matter (DM) are
\eq{\label{symms}
\begin{array}{lcl}
\mbox{orbifold:} & \rightarrow & x^{1} \leftrightarrow x^{2}~,~~x^3 \rightarrow - x^3 - N x^1 x^2~;\\
  \mbox{DM parity:} & \rightarrow & x^{1,2} \leftrightarrow -x^{1,2}~,~~x^3 \rightarrow x^3~. \end{array}
}
The wave-functions \eqref{uv} for $a=b=0$ can now be reorganised in terms of their parities under the orbifold projection and the DM parity: for the torus modes
\eq{
\begin{array}{c}
\mbox{orbifold even:}~~ \left\{ \begin{array}{l}
v_{l,n} + v_{n,l} + v_{-l,-n} + v_{-n, -l}\, ~~~ \mbox{DM parity even}\,, \\
v_{l,n} + v_{n,l} - v_{-l,-n} - v_{-n, -l}\, ~~~ \mbox{DM parity odd}\,, \end{array} \right.
\\
\mbox{orbifold odd:}~~ \left\{ \begin{array}{l}
v_{l,n} - v_{n,l} + v_{-l,-n} - v_{-n, -l}\, ~~~ \mbox{DM parity even}\,, \\
v_{l,n} - v_{n,l} - v_{-l,-n} + v_{-n, -l}\, ~~~ \mbox{DM parity odd}\,, \end{array} \right.
\end{array}
}
where $l \geq | n | \geq 0$ to avoid double counting; for the fiber modes
\eq{\hspace{-0.2in}
\begin{array}{c}
\mbox{orbifold even:}~~ \left\{ \begin{array}{l}
u_{k,l,n} + u_{-k,k-l,n} + (-1)^n ( u_{-k,l,n} + u_{k,k-l, n} )\, ~~~ \mbox{DM parity even}\,, \\
u_{k,l,n} - u_{-k,k-l,n} + (-1)^n ( u_{-k,l,n} - u_{k,k-l, n} )\, ~~~ \mbox{DM parity odd}\,, \end{array} \right.
\\
\mbox{orbifold odd:}~~ \left\{ \begin{array}{l}
u_{k,l,n} - u_{-k,k-l,n} + (-1)^n ( - u_{-k,l,n} + u_{k,k-l, n} )\, ~~~ \mbox{DM parity even}\,, \\
u_{k,l,n} + u_{-k,k-l,n} + (-1)^n ( - u_{-k,l,n} - u_{k,k-l, n} )\, ~~~ \mbox{DM parity odd}\,, \end{array} \right.
\end{array}
}
where $k>0$, and $0<l<k/2$ for even $k$ or $0<l<(k+1)/2$ for odd $k$, to avoid double counting.

\begin{table}[t] \begin{center}
\begin{tabular}{|c|c||cc||cc|}
\hline
   &  \multirow{2}{*}{$M_{KK}^2 \left(\frac{r}{2\pi} \right)^2$} & \multicolumn{2}{c||}{orbifold even} & \multicolumn{2}{c|}{orbifold odd} \\
 &  & DM-even & DM-odd & DM-even & DM-odd \\
\hline \hline
$(l,n)$ & \multicolumn{5}{c|}{Torus modes} \\
\hline
$(0,0)$                    & $0$            & 1 & - & - & - \\
$(l,0)$                     & $l^2$         & 1 & 1 & 1 & 1 \\
$(l,l)$ \& $(l,-l)$       & $2 l^2$      & 1 & 1 & 1 & 1 \\
$(l,|n|)$ \& $(l,-|n|)$ & $l^2+n^2$ & 2 & 2 & 2 & 2 \\
\hline \hline
$(k,n)$ & \multicolumn{5}{c|}{Fiber modes} \\
\hline
even $k$   &  \multirow{2}{*}{$|N| \left( \frac{k (2n+1)}{2\pi} + k^2 \xi\right)$} & $k/2$ & $k/2$ & $k/2$ & $k/2$ \\
odd $k$ & & $(k+1)/2$ & $(k-1)/2$  & $(k+1)/2$ & $(k-1)/2$  \\
\hline
\end{tabular}
\caption{Spectrum of a scalar field on the orbifold. For both odd and even cases, the tiers are labelled in terms of the DM parity. In the last four columns we report the degeneracy of each mass tier in both cases. We take here $k>0$.}\label{tab:spectrum} \end{center}
\end{table}

\begin{figure}[tb!]
\begin{center}
\includegraphics[width=10cm]{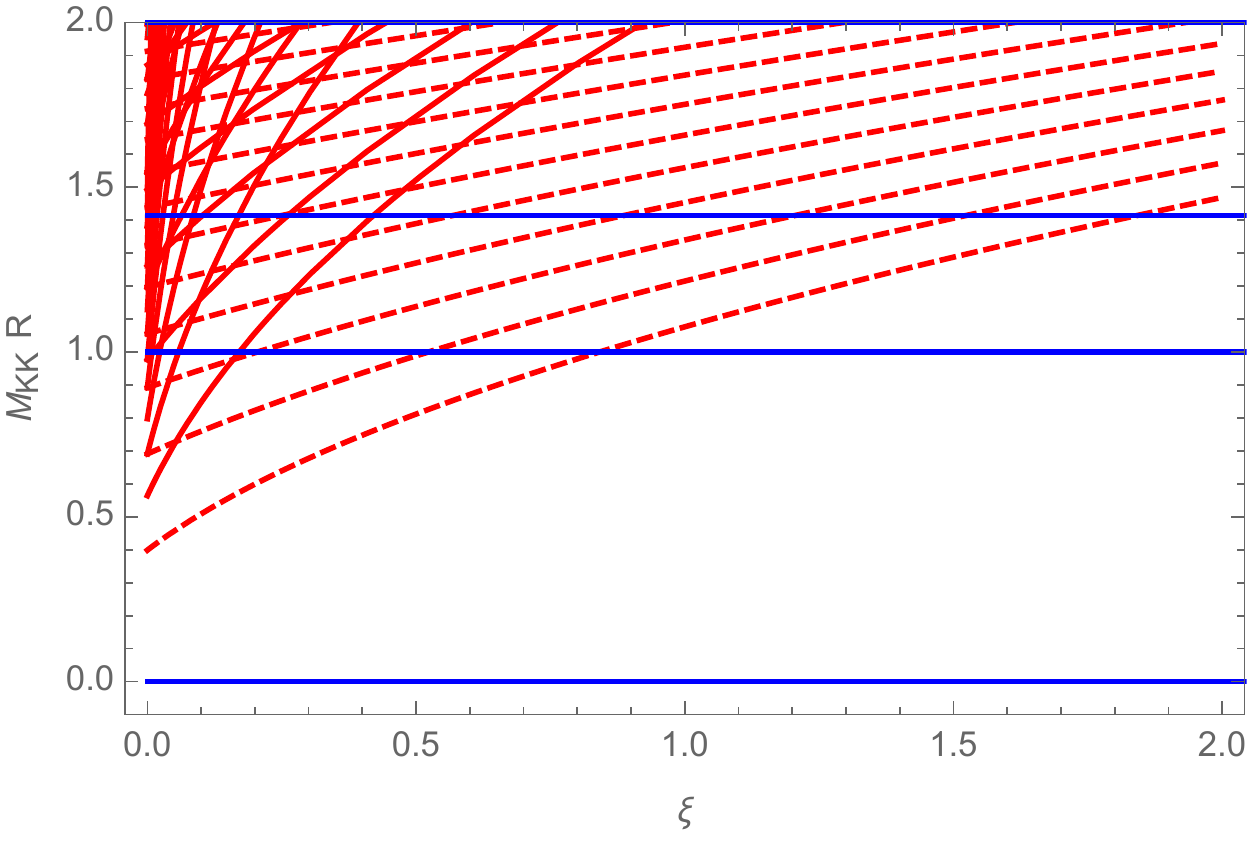}
\end{center}
\caption{Plot of the spectrum of a scalar field for $N=1$, projected on the orbifold-even states, as a function of $\xi$: in blue the lightest torus states, in red the fiber ones. The dotted red lines correspond to the fiber modes with $k=1$ that have no DM-odd states, while all the solid lines contain both DM-even and DM-odd states, with degeneracies as in Table~\ref{tab:spectrum}.} \label{fig:spectrum}
\end{figure}

A summary of the spectrum (for both orbifold even and odd scalar fields) is presented in Table~\ref{tab:spectrum}. The masses are expressed in units of the radius of the torus, $r/(2\pi)$, and we define a dimensionless parameter
\eq{
\xi = \frac{1}{|N|} \left(\frac{r}{c r^3}\right)^2}
that encodes the size of the third space dimension.
We note that all mass levels contain both DM-even and DM-odd states, except for the zero mode on the torus, and the modes with $k=1$ on the fiber. Furthermore, the lightest KK mode is the DM-even fiber mode $(k=1,l=0,n=0)$ for
\eq{
\xi < \frac{2 \pi - |N|}{2 \pi |N|} \left( = 0.84~~\mbox{for}~N=1\right)\,.}
For larger values, it is the torus mode $(1,0)$ that is the lightest.
Also, the DM state lives in the DM-odd $(k=2,l=0,n=0)$ fiber tier for
\eq{
\xi < \frac{\pi - |N|}{4 \pi |N|} \left( = 0.17~~\mbox{for}~N=1\right)\,,}
else it is in the lightest torus mode.
A plot of the spectrum for $N=1$ as a function of $\xi$ can be seen in Figure~\ref{fig:spectrum}: we clearly see that for small $\xi$, i.e. large $r^3$, a dense spectrum of fiber states forms above a mass gap determined by the radius of the torus base. For increasing $\xi$ (decreasing $r^3$), these states are lifted and for $\xi > 1$ the phenomenology is dominated by the torus modes alone.

\section{Conclusions and outlook} \label{sec:concl}

Nilmanifolds are a class of negatively-curved manifolds which offer the possibility to analytically calculate the spectrum of propagating fields.
This property can be very useful for the construction of effective models of new physics beyond the Standard Model. One attractive feature is that the spectrum of masses is quite different from that of the more familiar cases of flat or positively-curved spaces. Moreover these models are likely to be embeddable in string theory compactifications.

In this work, we considered three extra spatial dimensions, compactified as the three-dimensional Heisenberg nilmanifold. This space consists of a one-dimensional fiber over a two-dimensional torus base. After constructing an explicit metric and coordinate system on the manifold, we studied the eigenvalues and eigenfunctions of the three-dimensional Laplacian, which  directly determines the spectrum and wave-functions of a scalar field propagating in the bulk of the manifold. We found that the spectrum contains a complete tower of modes on the torus, which do not depend on the coordinate and radius of the fiber. Additionally, there are states whose masses only depend on the fiber radius and on the energy scale $\f$ related to the curvature. These fiber modes can be made lighter than the torus modes by tuning the various scales, as discussed in Section \ref{sec:DM}, and the energy gaps can be enhanced thanks to additional parameters in the general case \eqref{kgmcap}. The fiber modes are novel: the first term in their masses \eqref{masses} is the standard KK one, showing a mass gap given by the radius, but the second term involving also $\f$ is unusual and gives more finely-spaced modes following a linear Regge trajectory. The fiber can thus  provide a unique signature at a collider, if a particle physics model is built in this background.

As a first example, we built a model consisting of a singlet scalar field propagating in the bulk, while the SM fields are localised on a brane in four dimensions. The orbifolding of the nilmanifold is necessary in order to define fixed points where the SM brane might be localised. In the case we present, only fixed circles are possible, however the orbifold space has a residual isometry that can play the role of Dark Matter parity. This example shows that this class of models are indeed possible.

We discussed in the Introduction the problem of the low energy approximation in supergravity compactifications. The scalar spectrum obtained here does not allow to decouple the KK tower while keeping a few light massive modes. Indeed, despite the presence of the geometric flux $\f$ in the masses in addition to the radii, there is no approximation leaving a finite set of massive modes while getting rid of the tower: sending the radii to zero makes all fiber modes disappear and leaves only the torus base massless modes. This can also be seen when replacing $\frac{\f}{r_3}$ by $\frac{N}{r_1 r_2}$ in \eqref{masses}. If any, light massive modes (or equivalently masses for moduli) then do not come from the reduction of scalars, but from that of different fields.

The spectrum of fields carrying non-trivial spin is more complicated to obtain, but the algorithm presented and tested in Section \ref{sec:num} could be used for this purpose. This is the natural next step in order to allow the whole SM to propagate in the bulk of the nilmanifold instead of being localised. Indeed, while this paper focuses on a setup with a bulk scalar field whose excitations contain a dark matter candidate, it would be interesting to investigate the case in which all SM fields are bulk fields and the dark matter candidate is an excitation of a neutral SM field, as realized in other types of extra-dimensional models \cite{Maru:2009wu,Dohi:2010vc,Arbey,Agashe:2007jb,Ahmed:2015ona}. Furthermore, a more complete study of the orbifolds is necessary to identify a space that can accommodate both a Dark Matter candidate and chiral fermion zero modes. Our work is a first step in using negatively-curved spaces for particle phenomenology.

\section*{Acknowledgements}

The work of D.~A. is part of the Einstein Research Project ``Gravitation and High Energy Physics'', which is funded by the Einstein Foundation Berlin.
G.~C., A.~D., N.~D. and D.~T.  also acknowledge partial support from the D\'efiInphyNiTi - projet structurant TLF; the Labex-LIO (Lyon Institute of Origins) under grant ANR-10-LABX-66 and FRAMA (FR3127, F\'ed\'eration de Recherche ``Andr\'e Marie Amp\`ere'').

\begin{appendix}
\section{On the integrality of $N$}\label{sec:N}

\begin{figure}[tb!]
\begin{center}
\includegraphics[width=10cm]{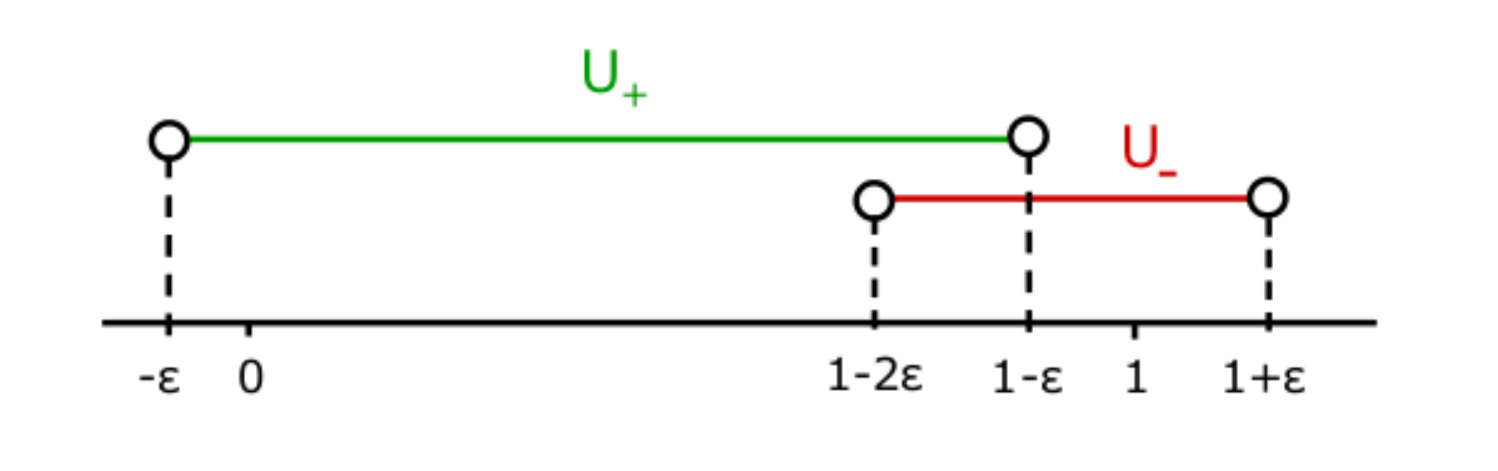}
\end{center}
\caption{A cover of $S^1_{x^1}$ consisting of the two open intervals
$U_{\pm}$.} \label{fig3}
\end{figure}

The integrality of $N$ given in \eqref{a} can be understood as follows. The nilmanifold is a circle fibration of an $S^1$ parameterised by $x^3$ fibered over a $T^2$ base parameterised by $x^1, x^2$.  Let us consider the associated principal $U(1)$ bundle with fiber parameterised by $\{g:=e^{2\pi \i x^3}, ~x^3\in[0,1]\}$.
From (\ref{a}) we see that the vertical displacement on the fiber can be rewritten as:
\eq{\left(\d x^3+N x^1\d x^2\right) = \frac{1}{2\pi \i}~\!g^{-1}\mathcal{D}g~,}
where $\mathcal{D}:=\d+A$ is the $U(1)$-covariant derivative with connection $A:=2\pi \i N  x^1\d x^2$. The base of the bundle may be covered by the
open patches $\mathcal{B}_+=U_+\times S^1_{x^2}$ and $\mathcal{B}_-=U_-\times S^1_{x^2}$, where $S^1_{x^m}$ denotes the circle parameterised by $x^m$; the open intervals $U_{\pm}$ furnish a cover of $S^1_{x^1}$ and are defined as:
\eq{U_+:=\{-\varepsilon< x^1< 1-\varepsilon\}~;~~~ U_-:=\{1-2\varepsilon< x^1< 1 +\varepsilon\}~,}
with $\varepsilon$ an infinitesimal positive number. An illustration is provided in Figure \ref{fig3}. The overlap $U_-\cap U_+$ consists of the point $\{x^1_+=0\}$, in terms of $U_+$ coordinates, or equivalently $\{ x^1_-=1\}$ in terms of $U_-$ coordinates.
Let us denote by $Q$  the point of $S^1_{x^1}$ parameterised by $x^1_+=0$, or equivalently $x^1_-=1$, i.e. $Q=U_-\cap U_+$.
The overlap $\mathcal{B}_-\cap \mathcal{B}_+= Q\times S^1_{x^2}$
is thus a copy of $S^1_{x^2}$. Let $t$ be the  transition function on $\mathcal{B}_-\cap \mathcal{B}_+$, so that $t~\!:~\!Q\times S^1_{x^2}\rightarrow  U(1)$.
Since $t$ is a map from $S^1_{x^2}$ to $S^1\cong U(1)$, it is
classified by $\pi_1(S^1)\cong\mathbb{Z}$ and we may set $t=e^{-2\pi \i M x^2}$, with $M\in \mathbb{Z}$.
The connection on $\mathcal{B}_{\pm}$ is given by $A_{\pm}=2\pi \i N  x_{\pm}^1\d x_{\pm}^2$ respectively.  On the overlap $\mathcal{B}_-\cap \mathcal{B}_+= Q\times S^1_{x^2}$,  these are related via:
\eq{\label{trans}A_+=t^{-1}A_-t+t^{-1}\d t~.}
Evaluating the above at $Q$ ($x^1_+=0$, $x^1_-=1$) we obtain $t=e^{-2\pi \i N x^2}$. For the latter to be a well-defined element of $U(1)$ for all $x^2\in S^1_{x^2}$, $N$ must be an integer. Alternatively we can arrive at the same conclusion by noting
that the first Chern class of the principal $U(1)$-bundle is integral in cohomology and is given by $c_1=N\d x^1\wedge\d x^2$.

To obtain the twist of the fiber coordinate, let
$x^3_{\pm}$ be the coordinate  of the $S^1_{x^3}$ fiber over $\mathcal{B}_{\pm}$ respectively and let
$g_{\pm}=e^{2\pi \i x_{\pm}^3}$ denote the corresponding points on the fiber of the associated principal $U(1)$ bundle.  On the overlap $\mathcal{B}_-\cap \mathcal{B}_+= Q\times S^1_{x^2}$  these are related via $g_-=t\cdot g_+$,
which leads to
\eq{\label{gluing}
x_{-}^3=x_{+}^3-N x^2
~.}
Note that the above equation is
invariant under $x^2\sim x^2+1$, since
$x^3$ itself is only defined modulo integral shifts and $N$ is an integer.

Finally, let us give a further derivation of these results. We first consider the lattice identifications \eqref{heisids} for $n^{m=1,2,3}$ being either zero or some fixed value in $\mathbb{Z}^*$, and $N\in \mathbb{R}^*$: those can be rewritten as
\eq{
(x^1,x^2,x^3) \equiv (x^1,x^2 + n^2,x^3) \equiv (x^1,x^2,x^3+ n^3) \equiv (x^1 + n^1,x^2,x^3-n^1 N x^2) \ ,
}
with $n^m \in \mathbb{Z}^*$. Using each of them, one can prove the following chains of identifications:
\eq{
(x^1,x^2,x^3) \equiv (x^1,x^2,x^3+ n^3) \equiv \dots \equiv (x^1,x^2,x^3+ N' n^3) \ ,
}
for any $N' \in \mathbb{Z}^*$ and
\eq{\spl{
(x^1,x^2,x^3) & \equiv (x^1,x^2 + n^2,x^3) \equiv (x^1 + n^1,x^2 + n^2,x^3-n^1 N n^2-n^1 N x^2) \\
 &\equiv (x^1 + n^1,x^2 ,x^3-n^1 N n^2-n^1 N x^2) \equiv (x^1,x^2 ,x^3-n^1 N n^2) \ .
}}
For consistency, one should have
\eq{
N' = \frac{ n^1 n^2}{n^3} N \in \mathbb{Z}^* \Leftrightarrow N = \frac{n^3}{ n^1 n^2} N' \in \mathbb{Q}^*  \ .
}
This reproduces for our $\mcal_3$ the mathematical result by Malcev \cite{Malcev}, stating that for nilpotent groups, a lattice exists (allowing to build the nilmanifold) if and only if the structure constants of the algebra are rational in some basis. After rescaling the algebra by the radii, $N$ is nothing but the non-zero structure constant, and we indeed conclude that having identifications by a lattice is equivalent to $N$ being rational. In addition, specialising to the case $n^m=1$, one deduces $N=N' \in \mathbb{Z}^*$, in agreement with the above result.

\section{Orthonormal modes}\label{ap:ortho}

In this appendix, we show that the modes \eqref{uultcap} are orthonormal, i.e.
\eq{\label{orthon}
\int \d^3 x \sqrt{g}\, U_{k,l,n}(x^1, x^2, x^3) U^*_{k',l',n'}(x^1, x^2, x^3) = \delta_{k,k'} \delta_{l,l'} \delta_{n,n'} \ .
}
To that end, we compute the left-hand side given by
\eq{\spl{
I= \frac{r^2}{|N|}\frac{1}{2^n n! \sqrt{\pi}} \sum_{m,m' \in \mathbb{Z}} \int_{[0,1]^3} \d^3 x\, & e^{2\pi \i x^3 (k-k')} e^{2\pi \i \f \frac{r^1}{r^3} x^1 (k z_{m,k,l} - k' z_{m',k',l'})}\\
\times &\, e^{-\frac{\i\pi b}{r^3(b^2+c^2)} \left(k z_{m,k,l} (\f a z_{m,k,l} - 2) - k' z_{m',k',l'} (\f a z_{m',k',l'} - 2)\right)}\\
\times &\, \Phi^{\lambda}_n(w_{m,k,l}) \Phi^{\lambda}_{n'}(w_{m',k',l'}) \ ,
}}
where $z_{m,k,l}$ and $w_{m,k,l}$ correspond to $z_m$ and $w_m$, and we refer to Section \ref{sec:specgen} for the definitions of the various terms. First, the integral over $x^3$ gives $\delta_{k,k'}$. Then, the integral over $x^1$ imposes similarly $l-l'=-k(m-m')$. Since $0\leq l \leq |k|-1$ and similarly for $l'$, one deduces $|m-m'|<1$ i.e. $m=m'$, thus $l=l'$. We deduce $z_{m,k,l} = z_{m',k',l'}$ and similarly for $w_m$. We are then left with
\eq{
I= \delta_{k,k'} \delta_{l,l'} \frac{r^2}{|N|}\frac{1}{2^n n! \sqrt{\pi}} \sum_{m \in \mathbb{Z}} \int_0^1 \d x^2\,  \Phi^{\lambda}_n(w_{m,k,l}) \Phi^{\lambda}_{n'}(w_{m,k,l}) \ .
}
We recall that $ w_{m,k,l}=\frac{r^2}{N} \left(N x^2+m+\omega\right)$ with $\omega=\frac{l}{k} -\frac{aN}{r^2\f(a^2+b^2+c^2)}$. So
\eq{\spl{
I & = \delta_{k,k'} \delta_{l,l'} \frac{r^2}{|N|}\frac{1}{2^n n! \sqrt{\pi}} \sum_{m \in \mathbb{Z}} \int_m^{N+m} \frac{\d y}{N}\,  \Phi^{\lambda}_n\left(\frac{r^2}{N} \left(y+\omega\right)\right) \Phi^{\lambda}_{n'}(\dots) \\
& = \delta_{k,k'} \delta_{l,l'} \frac{r^2}{2^n n! \sqrt{\pi}} \sum_{m \in \mathbb{Z}} \int_m^{m+1} \frac{\d y}{|N|}\,  \Phi^{\lambda}_n\left(\frac{r^2}{N} \left(y+\omega\right)\right) \Phi^{\lambda}_{n'}(\dots) \\
& = \delta_{k,k'} \delta_{l,l'} \frac{r^2}{2^n n! \sqrt{\pi}} \int_{-\infty}^{+\infty} \frac{\d y}{|N|}\,  \Phi^{\lambda}_n\left(\frac{r^2}{N} \left(y+\omega\right)\right) \Phi^{\lambda}_{n'}(\dots) \\
& = \delta_{k,k'} \delta_{l,l'} \frac{1}{2^n n! \sqrt{\pi}} \int_{-\infty}^{+\infty} \d z\, \Phi^{\lambda}_n(z) \Phi^{\lambda}_{n'}(z) \\
& = \delta_{k,k'} \delta_{l,l'} \frac{1}{2^n n! \sqrt{\pi}} \int_{-\infty}^{+\infty} \d u\, \Phi_n(u) \Phi_{n'}(u) \\
& = \delta_{k,k'} \delta_{l,l'} \frac{1}{2^n n! \sqrt{\pi}} \int_{-\infty}^{+\infty} \d u\, e^{-u^2} H_n(u) H_{n'}(u) \\
& = \delta_{k,k'} \delta_{l,l'}  \delta_{n,n'} \ .
}}
This concludes our proof of \eqref{orthon}.

As a side remark, consider a smooth eigenfunction $U$ of eigenvalue $\lambda_0$. Let us use for $U$ the same norm as in \eqref{orthon} and take it to be non-zero; we define with the metric an analogous norm for a one-form. Then, we have on the compact manifold (without singularity)
\eq{\label{0mode}\spl{
0 \leq ||\d U ||^2 & = \int \d^3 x\, \sqrt{g}\, g^{mn} \del_m U^* \del_n U = 0 - \int \d^3 x\, U^*  \del_m \left(\sqrt{g}\, g^{mn} \del_n U \right)\\
& =  - \int \d^3 x\, \sqrt{g}\, U^* \Delta U =  - \lambda_0 || U ||^2 \ .
}}
We deduce that $\lambda_0 \leq 0$ and $\lambda_0=0 \leftrightarrow \d U =0$. This means that there is no tachyon, and the only massless modes are constant functions. Consequently, modes depending on $x^3$ cannot be massless. It should be possible to extend this reasoning to differential forms, with closed and co-closed forms.

\section{On the completeness of the set of modes}\label{ap:complete}

In this appendix we argue that the set of Laplacian eigenmodes found in the main text, namely $U_{k,l,n}$ in \eqref{uultcap} and $V_{p,q}$ in \eqref{52}, is complete. To show this, we verify that the most general normalisable solutions to the differential equation have been found, given the boundary conditions. To that end, we first give the most general form of the functions satisfying the boundary conditions, namely the identifications \eqref{heisidscap}, and then solve the equation for those. For eigenfunctions independent of $X^3$, \eqref{heisidscap} simply indicates functions periodic in $X^1$ and $X^2$. Such a function can be written in full generality as two Fourier series, leading to the modes $V_{p,q}$. Then, solving the equation does not introduce new constraints. So we turn to the case of a non-trivial dependence on $X^3$.

\subsection*{Boundary conditions}

Consider a function $U(X^1,X^2,X^3)$ that satisfies the boundary conditions \eqref{heisidscap} with $n^1=n^2=n^3=1$. First of all, it is periodic in $X^3$ of period $r^3$. It can thus be written in full generality as a Fourier transform
\eq{
U(X^1,X^2,X^3)= \sum_{k\in \mathbb{ Z}} e^{2\pi \i K X^3} c_k(X^1, X^2) \ ,
}
where we use the notation \eqref{la}. As we are interested in a dependence on $X^3$, we focus on the modes with $K\neq0$. For convenience, we rewrite $c_k$ as
\eq{
c_k(X^1, X^2)=d_{k,l}(X^1, X^2) e^{2\pi \i (K \f X^1 X^2 + L X^1)}\ .
}
We now study the boundary condition $X^1 \rightarrow X^1 +r^1,\ X^3\rightarrow X^3 - \f r^1 X^2$: identifying each $X^3$ mode, we arrive at the condition
\eq{
d_{k,l}(X^1 +r^1, X^2) = d_{k,l}(X^1, X^2) \ . \label{cond1}
}
In addition, $U$ should as well be periodic under $X^2 \rightarrow X^2 +r^2$, last of the three boundary conditions. This translates into
\eq{
d_{k,l}(X^1, X^2 +r^2)  = e^{-2\pi \i K \f X^1 r^2 } d_{k,l}(X^1, X^2) \label{cond2} \ .
}
The periodicity condition \eqref{cond1} could lead to a Fourier series with coefficients depending on $X^2$. The remaining condition \eqref{cond2} translated on these coefficients is then not easy to solve. We proceed differently and introduce the Zak transform $Zf (w,t)$ of a function $f(t)$, defined as
\eq{
Zf (w,t) = \sum_{m\in \mathbb{ Z}} e^{2\pi \i mw} f(t+m) \ , \label{Zak}
}
given some conditions on $f$ that we will come back to. This transform verifies precisely the two properties of periodicity \eqref{cond1} and translation \eqref{cond2}, up to appropriate normalisations. In addition, this transformation is invertible. We thus consider that $d_{k,l}$ is the Zak transform of a function $f_{k,l}(X^2)$. Following \cite{Brezin, thang}, this should be the only solution to the boundary conditions. We get
\eq{
d_{k,l}(X^1, X^2) = \sum_{m\in \mathbb{ Z}} e^{2\pi \i M K X^1} f_{k,l}\left(X^2+\frac{M}{\f}\right) \ ,
}
and one can verify that \eqref{cond1} and \eqref{cond2} are satisfied using \eqref{la} and \eqref{quant}. We conclude that the modes $U_{k,l}$ in \eqref{wbzcap} with $k\neq 0$ are the most general ones verifying the boundary conditions \eqref{heisidscap}.

\subsection*{Solving the equation}

We turn to the differential equation and follow the procedure presented in Section \ref{sec:specgen}. Going from $F$ to $G$ to $H$ to $T$ are smooth operations that can be done in full generality. Furthermore we  perform the redefinitions $T(w_m)=|\lambda|^{\frac{1}{4}} \Phi (|\lambda|^{\frac{1}{2}} w_m)= |\lambda|^{\frac{1}{4}} e^{-\frac{|\lambda|}{2} w_m^2} {\cal H} (|\lambda|^{\frac{1}{2}} w_m)$ where $\Phi$ and ${\cal H}$ are for now completely general. We redefine $y=|\lambda|^{\frac{1}{2}} w_m$ and choose $\lambda$ as in \eqref{lambda}. Our initial equation $(\nabla^2 + M^2) U_{k,l}=0$ becomes (identifying the $X^1$ or $m$ modes)
\eq{\spl{\label{Hermite}
& \del_y^2 {\cal H} - 2 y \del_y {\cal H} + \Lambda {\cal H} = 0 \ ,\\
& \mbox{with}\ \Lambda=\frac{1}{|\lambda|} \left(\frac{c^2}{b^2+c^2} M^2 - |\lambda| - A \right) \ , \quad A= \frac{4\pi^2K^2c^2}{(b^2+c^2)(a^2+b^2+c^2)} \ .
}}
For a generic $\Lambda \in \mathbb{R}$, this is precisely the Hermite differential equation. Using series, this equation can be shown to admit two independent solutions, sometimes called confluent hypergeometric functions of the first kind. These series solutions, one of which consists of odd powers of $y$ and the other of even powers, converge for all $y$ so are defined without restrictions (a particular case of Fuch's theorem). For $\Lambda=2n \geq 0$ and for these values only, one of these two series (depending on $n$ being an even or odd integer) gets truncated to a Hermite polynomial; the other one remains an infinite series. In Sections \ref{sec:simple} and \ref{sec:specgen}, we took precisely this $\Lambda$ and the Hermite polynomial to solve the equation. More generally for $\Lambda \in \mathbb{R}$, one would obtain the following mass
\eq{\label{mass}
M^2= \left(A + |\lambda| (1 + \Lambda) \right) \frac{b^2+c^2}{c^2} \ .
}
Such a continuous spectrum is not consistent with the fact that the Laplacian spectrum of normalisable functions on a compact manifold should be discrete. Here, the only distinction among solutions allowing a discretisation would be the value $\Lambda=2n \geq 0$ and the Hermite polynomials. One may then wonder about the infinite series solutions. They turn out to be very divergent at infinity. For instance, when $\Lambda \neq 2n$, the even series behave at $y \sim \infty$ as a constant times $e^{y^2}$. Coming back to $T$, the $e^{-\frac{1}{2}y^2}$ factor is not enough to compensate this divergence. Coming back to $F$, one obtains at $X \sim \infty$
\eq{
|F(X)| \sim e^{\frac{|\lambda|}{2}(X)^2} \ .
}
As a consequence, the Zak transform on $F$ cannot be defined (its defining series does not converge). More precisely $F\notin {\cal L}^1 (\mathbb{R}, \mathbb{C}),\ F\notin {\cal L}^2 (\mathbb{R}, \mathbb{C})$, nor does $F$ satisfy the ``decay condition'', which is problematic for the Zak transform as discussed in chapter 16 of \cite{Zakbook}. This illustrates why the infinite series solutions and their continuous spectrum are excluded. On the contrary, for a (Hermite) polynomial ${\cal H}(y)$, the combination with $e^{-\frac{1}{2}y^2}$ in $T$ is not divergent at infinity, and the convergence is good enough to define the Zak transform.

We conclude that the modes $U_{k,l,n}$ in \eqref{uultcap} are the only solutions to our problem. They are normalisable wave-functions, as shown in Appendix \ref{ap:ortho}, and give the discrete spectrum \eqref{kgmcap}.

\end{appendix}

\end{document}